\documentclass[aps,amsmath,amssymb,prd,showpacs,floatfix,preprint,superscriptaddress,nofootinbib,12pt]{article}
\usepackage{jheppub}
\usepackage{bm}
\usepackage{ulem}

\usepackage{axodraw4j}
\usepackage{pstricks}
\usepackage{color}

\allowdisplaybreaks[3]

\begin{document}

\title{Long-range fermions and critical dualities}

\author[a]{Noam Chai}
\author[b]{Soumangsu Chakraborty}
\author[c]{Mikhail Goykhman}
\author[a]{Ritam Sinha}
\affiliation[a]{The Racah Institute of Physics, The Hebrew University of Jerusalem, \\ Jerusalem 91904, Israel}
\affiliation[b]{Department of Theoretical Physics, Tata Institute of Fundamental Research,
Homi Bhabha Road, Mumbai 400005, India}
\affiliation[c]{William I. Fine Theoretical Physics Institute, University of Minnesota, Minneapolis, MN
55455, USA}
\emailAdd{ritam.sinha@mail.huji.ac.il}
\emailAdd{goykhman@umn.edu}
\emailAdd{soumangsuchakraborty@gmail.com}
\emailAdd{ritam.sinha@mail.huji.ac.il}

\abstract{
We construct long-range fermionic models with the Gross-Neveu and Gross-Neveu-Yukawa
interaction, and argue that their critical regimes are equivalent. To this end, we calculate
various CFT data in $\epsilon$- and $1/N$- expansion, and demonstrate their agreement
in the overlapping regimes of validity.
}

\maketitle

\section{Introduction}
\label{sec:intro}

The Gross-Neveu (GN) model describes dynamics of fermions in fundamental
representation of the $U(n)$ symmetry group, coupled via the four-fermionic
interaction term \cite{Gross:1974jv}. In two dimensions, this model
provides a toy example of quantum chromodynamics, featuring
such properties as asymptotic freedom, dynamical chiral symmetry breaking, and mass gap generation via dimensional
transmutation. In $2<d<4$ dimensions
the four-fermionic interaction is power-counting non-renormalizable.
However, the Gross-Neveu model is renormalizable when expanded in $1/n$
for a large number $n$ of fermionic flavors \cite{Gross:1974jv,Zinn-Justin:1991ksq}. 

In fact, for $2<d<4$, renormalization group (RG) flow of the four-fermionic coupling constant brings the model
to a fixed point in the ultra-violet (UV) limit, manifesting a non-trivial interacting critical regime at high energies.
Critical regimes at fixed points of RG flows can be studied via
techniques of conformal field theory (CFT) \cite{Polyakov:1970xd}.\footnote{In a weakly-coupled regime,
one can explicitly solve for the fixed point values of coupling constants using
perturbative $\epsilon$-expansion \cite{Wilson:1971dc}. In a general non-perturbative regime, but at large $N$,
existence of a non-trivial CFT
at the end of an RG flow in vector models can be  argued for using the Hubbard-Stratonovich
transformation \cite{Gubser:2002vv}. By demanding that the Hubbard-Stratonovich field
obeys the unitarity bound, one 
constrains the allowed space-time dimension \cite{Parisi:1975im}.
In case of the Gross-Neveu model, this leads to the requirement $d<4$ \cite{Gross:1974jv,Zinn-Justin:1991ksq}.}
The same critical regime has been argued
to be attained at the infra-red (IR) end of an RG flow of a different system, the
Gross-Neveu-Yukawa (GNY) model  \cite{Zinn-Justin:1991ksq}. This equivalence of two
different models at criticality therefore furnishes an example of a critical universality.
Since one of these models achieves its interacting critical regime in the UV, while another one is taken
in the IR, one can in fact view this critical universality as a UV completion.

A bosonic model counterpart of such a UV completion has recently been argued
to exist for the $O(N)$ vector model with quartic interaction \cite{Fei:2014yja}. When considered in
higher dimensions, $4<d<6$, the quartic scalar coupling is power-counting non-renormalizable.
However, critical regime of the large-$N$ version of this model, achieved in the UV, can be explored in the $1/N$
expansion \cite{Parisi:1975im}. A UV completion of this model was proposed in \cite{Fei:2014yja}, where it was
argued that the same CFT can be found at the IR end of an RG flow of the 
$O(N)$ vector model coupled to a fundamental singlet scalar field with a cubic self-interaction.
When considered in overlapping regimes
of validity, the corresponding CFT data were shown to match for these two models \cite{Fei:2014yja,Fei:2014xta,Gracey:2015tta}.\\

A different class of critical models is given by spin systems with long-range interaction,
and their continuous version, provided by generalized free scalar field perturbed by a local quartic interaction
term \cite{Dyson:1968up,Fisher:1972zz,Kosterlitz:1976,Aizenman1988}. These models
are characterized by an extra parameter $s$, controlling the exponent of the power-law
decay of the long-range bi-local kinetic term of the generalized free field.
When $s>d/2$, the quartic coupling constant is relevant, and
 these models reach a non-trivial critical regime in the IR (when $s<d/2$,
the IR regime is described by the mean-field theory).
The $O(N)$ vector model version of
these models has been explored in \cite{Gubser:2017vgc,Giombi:2019enr,Chai:2021arp}.

For $s$ greater than the
crossover value, $s>s_\star = 2-2\gamma_{\hat\phi}$, where $\gamma_{\hat\phi}$
is anomalous dimension of the short-range fundamental scalar field $\hat\phi$,  the model smoothly
transitions to the short-range critical regime \cite{Sak:1973,Sak:1977}. 
The nature of this short-range model has been recently elucidated in \cite{Behan:2017dwr,Behan:2017emf}.
It is given by the critical short-range model with quartic self-interaction plus a decoupled
generalized free field $\chi$ with scaling dimension $(d+s)/2$.

On the other hand, in the long-range domain $d/2<s<s_\star$, $1<d<4$,
one obtains a rich critical IR behavior, controlled by the parameter $s$. While long-range models
do not possess a stress-energy tensor, substantial evidence has been accumulated that this interacting critical regime
is in fact described by a CFT \cite{Paulos:2015jfa,Behan:2017dwr,Behan:2017emf,Behan:2018hfx} (see also \cite{Chai:2021arp} for the
investigation of conformal symmetry in the large-$N$ long-range $O(N)$ vector model, and \cite{Brydges:2002wq,Abdesselam:2006qg,Brezin2014,Slade:2016yer,Gubser:2017vgc,Giombi:2019enr,Benedetti:2020rrq} for previous calculations of critical exponents).
Moreover, the same
CFT was argued in \cite{Behan:2017dwr,Behan:2017emf} to be the IR end-point of an RG flow triggered by coupling
the short-range vector field $\hat\phi$ to the generalized free field $\chi$ of dimension $(d+s)/2$
via the bi-linear term $\lambda\,\int d^dx \hat\phi\chi$. When $s>s_\star$,
the coupling $\lambda$ is irrelevant, and the field $\chi$ decouples. For $s$ slightly below
$s_\star$, the IR fixed point $\lambda_\star$ can be studied perturbatively. $O(N)$
generalization of this duality has recently been discussed in \cite{Chai:2021arp}.

In the regime of $s<d/2$, the quartic interaction of the long-range vector model becomes irrelevant.
Therefore the IR regime of the model is described by mean-field theory.
However, RG flow of the quartic coupling drives
the theory to a non-trivial regime in the UV. The corresponding CFT can be studied
using the $1/N$ expansion. In fact, the results of \cite{Gubser:2017vgc,Giombi:2019enr,Chai:2021arp}
for various anomalous dimensions and OPE coefficients, derived for general $d$ and $s$,
can be formally continued to this regime. It was argued in  \cite{Chakraborty:2021lwl}
that for $s<d/2$ the long-range $O(N)$
vector model achieves a non-trivial critical regime in the UV. This situation
is analogous to a non-trivial UV criticality of the usual local (short-range) $O(N)$
vector model in $4<d<6$.
Having this analogy in mind, \cite{Chakraborty:2021lwl}
proposed a long-range version of the cubic model of \cite{Fei:2014yja} and provided 
perturbative evidence that the IR stable fixed point of this cubic model is equivalent
to the UV fixed point of the long-range $O(N)$ vector model for $d/3 < s < \min (d/2, s_\star)$, $1<d<6$.\\

In this paper, we are interested in constructing long-range version of the Gross-Neveu
model. We will begin by formulating a generalized free field theory action
for the Dirac fermion multiplet $\psi^i$, $i=1,\dots,n$ in fundamental representation of the $U(n)$ symmetry group.
Power-law decay of the long-range interaction will be controlled by an exponent parameter $s$.
We will then perturb this model by the local four-fermionic
self-interaction of the Gross-Neveu type, and argue that the resulting system
flows to an interacting regime in the UV in the domain\footnote{See also \cite{Giuliani:2020aot} for a similar
model involving long-range fermions.}
\begin{equation}
\label{fermion duality parameters}
d/2<s<\min(d/2+1,s_\star)\,,\qquad 1<d<4\,.
\end{equation}
Here $s_\star = 2-2\gamma_{\hat\psi}$ is the crossover value of the  exponent $s$,
above which the model transitions to the critical short-range regime, involving dynamics
of the interacting fermionic multiplet $\hat\psi^i$, $i=1,\dots,n$ with the anomalous
dimension $\gamma_{\hat\psi}$.

Working within the Hubbard-Stratonovich framework,
we will calculate various CFT data in the long-range Gross-Neveu model,
as a function of general $d$, $s$, at the next-to-leading 
order in $1/N$ expansion. In particular, we are interested in anomalous dimension
of the Hubbard-Stratonovich field, $\sigma$, and the OPE coefficient $\langle\sigma\bar\psi\psi\rangle$.
We will explicitly verify that all of our CFT data is in fact continuous
at the long-range--short-range crossover point $s_\star$.

As we have reviewed above, the short-range Gross-Neveu-Yukawa model furnishes
a UV completion of the short-range Gross-Neveu model \cite{Zinn-Justin:1991ksq}.
Motivated by this, and encouraged by the recent example \cite{Chakraborty:2021lwl}
of critical universality in the long-range models, we proceed to constructing
the long-range version of the Gross-Neveu-Yukawa model. Working within the scope
of perturbative $\epsilon$-expansion, we derive the IR stable fixed point of this model,
and calculate various CFT data at the fixed point.
Assembling all the results together, and expanding them in overlapping regimes
of validity, we explicitly demonstrate critical equivalence of the
long-range Gross-Neveu and Gross-Neveu-Yukawa models. We then suggest that the
critical duality in fact holds in the entire domain (\ref{fermion duality parameters}).\\

The rest of this paper is organized as follows.
In section~\ref{sec: SR GN} we collect some known CFT data in the
short-range Gross-Neveu model. 
In section~\ref{sec:LR GN} we construct the long-range Gross-Neveu model, and establish the
parameter domain (\ref{fermion duality parameters}) in which it flows to an interacting critical regime in the UV.
Working at the next-to-leading order in the $1/N$ expansion, we derive various CFT data.
We also explicitly demonstrate continuity of
CFT data at the long-range--short-range crossover point $s_\star$, using various
expressions for the short-range model, collected in section~\ref{sec: SR GN}.
In section~\ref{sec: LR GNY} we construct the long-range Gross-Neveu-Yukawa model.
Working perturbatively at the next-to-leading order in couplings, we find the IR stable fixed point
of this model, and calculate anomalous dimension of the scalar field at the fixed point.
We perform comparison of various CFT data in the long-range Gross-Neveu and
Gross-Neveu-Yukawa models in the overlapping regimes of validity, and demonstrate their agreement.
We discuss our results in section~\ref{sec:discussion}. Some known identities,
useful for CFT calculations in position space, are collected in appendix~\ref{app_a}.

\section{Brief review of short-range Gross-Neveu model}
\label{sec: SR GN}
As we discussed in section~\ref{sec:intro}, we expect to observe
a smooth crossover between the long-range and the short-range critical $U(n)$
fermionic models with the Gross-Neveu interaction
at the threshold value $s_\star$ of the long-range exponent parameter $s$. This threshold value can be expressed
 in terms
of anomalous dimension of the short-range fermion $\hat\psi$ as $s_\star = 2 - 2\gamma_{\hat\psi}$. In the large-$N$
limit, that we are going to be mostly interested in when studying Gross-Neveu interaction, the
fermion anomalous dimension is given by \cite{Zinn-Justin:1991ksq,Gracey:1990wi}
\begin{equation}
\label{gamma hat psi}
\gamma_{\hat\psi}=-\frac{1}{N}\frac{2^{d-1} \sin \left(\frac{\pi  d}{2}\right)
\Gamma \left(\frac{d-1}{2}\right)}{\pi ^{3/2} d \Gamma \left(\frac{d}{2}-1\right)}
+{\cal O}\left(\frac{1}{N^2}\right)\,.
\end{equation}
To facilitate test of smooth CFT crossover, in this section we collect some known
CFT data in the short-range Gross-Neveu model \cite{Gross:1974jv} \footnote{Euclidean gamma-matrices are Hermitian, $(\gamma^\mu)^\dagger = \gamma^\mu$,
satisfying $\gamma^\mu\gamma^\nu+\gamma^\nu\gamma^\mu =2\,\delta^{\mu\nu}\,\mathbb{I}$, where $\mathbb{I}$
is rank-$2^{[d/2]}$ unit matrix.
Dirac conjugate of a spinor is an ordinary Hermitian conjugate, $\bar\psi = \psi^\dagger$. Following the standard conventions in large-$N$ GN model \cite{Zinn-Justin:1991ksq}, we defined
$N=2^{[d/2]}\,n$.}
\begin{align}
\label{GN action}
S_{\textrm{SR\;GN}} = \int d^dx \, \left(  \bar{\hat\psi}^i \gamma^\mu \partial _\mu \hat\psi ^i +
\frac{\hat g}{N}\,\left(\bar{\hat\psi}\hat\psi)^2 \right)\right)\,,
\end{align}
where we a sum over repeated $U(n)$ index $i=1,\dots,n$ is implied.
Performing the Hubbard-Stratonovich transformation, we can 
re-write (\ref{GN action}) as (we skip $U(n)$ indices in what follows for the sake of brevity)
\begin{align}
\label{GN action HS}
S_{\textrm{SR\;GN}}  = \int d^dx \, \left( \bar{\hat\psi} \gamma^\mu \partial _\mu \hat\psi 
-\frac{1}{4 \hat g}\,\hat\sigma^2 + \frac{1}{\sqrt{N}}\,\hat\sigma\bar{\hat\psi}\hat\psi \right)\,.
\end{align}
While at the free fixed point, $\hat g=0$,
scaling dimension of the pseudo-scalar Hubbard-Stratonovich field $\hat\sigma\sim \bar{\hat\psi}\hat\psi$ 
is given by $[\hat\sigma]_{\textrm{free}} = d-1$, at the interacting fixed point, achieved in the UV
for $2<d<4$, it is given by $[\hat\sigma]_{\textrm{UV}} = 1 + \gamma_{\hat\sigma}$, where \cite{Zinn-Justin:1991ksq}
\begin{equation}
\label{gamma hat sigma}
\gamma_{\hat\sigma} \,=\, \frac{1}{N}\frac{4 \sin \left(\frac{\pi  d}{2}\right) \Gamma (d)}{\pi  d \Gamma \left(\frac{d}{2}\right)^2}
+{\cal O}\left(\frac{1}{N^2}\right)\,.
\end{equation}
Besides the scaling dimensions (\ref{gamma hat psi}), (\ref{gamma hat sigma}),
another non-trivial piece of CFT data is given by the amplitude of the 
three-point function \cite{Goykhman:2020ffn}\footnote{See also \cite{Manashov:2017rrx}
for earlier derivation of the $\hat\sigma\hat{\bar\psi}\hat\psi$ conformal triangle and $1/N$
corrections to the $\langle\hat\sigma\hat\sigma\rangle$ propagator.}
\begin{equation}
\label{psi psi s GN}
\begin{aligned}
&\langle \bar{\hat\psi}(x_1)\hat\psi(x_2) \hat\sigma(x_3)\rangle\Bigg|_{\textrm{normalized}} \,=\, 
C_{\bar{\hat\psi}\hat\psi \hat\sigma}\,(1+\delta C_{\bar{\hat\psi}\hat\psi \hat\sigma})\,\frac{\gamma_\mu x_{13}^\mu
\gamma_\nu x_{32}^\nu}{|x_{12}|^{d-2}|x_{13}|^{2}|x_{23}|^{2}}\,,\\
&C_{\bar{\hat\psi}\hat\psi\hat \sigma}\,=\,-\frac{2^\frac{d}{2}}{\sqrt{N}(d-2)\pi^\frac{3}{4}}\,
\sqrt{-\frac{\sin \left(\frac{\pi  d}{2}\right) \Gamma \left(\frac{d-1}{2}\right)}{\Gamma \left(\frac{d}{2}-1\right)}}\,,\\
&\delta C_{\bar{\hat\psi}\hat\psi\hat \sigma}\,
=\,\frac{2 (d-1) \left((d-2) H_{d-2}+\pi  (d-2) \cot \left(\frac{\pi  d}{2}\right)-2\right)}{(d-2)^2}\,
\gamma_{\hat\psi}\,,
\end{aligned}
\end{equation}
where propagators of all fields in position space have been normlalized to unity, and $H_n$ is the $n$th
Harmonic number.

\section{Long-range Gross-Neveu model}
\label{sec:LR GN}

Consider generalized (long-range) free Dirac fermion multiplet in fundamental
representation of $U(n)$, defined in $d$-dimensional Euclidean space, and
described by the bi-local action
\begin{equation}
\label{free generalized psi action}
S_{\textrm{LR\;Dirac}}[\psi] =C_f(s)\, \int d^dx\int d^dy\,\bar\psi^i(x)\gamma^\mu\psi^i(y)
\,\frac{(x-y)^\mu}{|x-y|^{d+s}}\,,
\end{equation}
where sum over repeated $i=1,\dots,n$ is implied.
Here $s$ is a free parameter, that determines the fermion scaling dimension
\begin{equation}
\label{definition of Delta_psi}
\Delta_\psi = \frac{d-s+1}{2}\,.
\end{equation}
We choose the pre-factor $C_f(s)$ so that the free fermionic propagator is canonically
normalized in momentum space,
\begin{equation}
\langle \psi(k)\bar\psi(p)\rangle = (2\pi)^d \delta^{(d)}(p+q)\,\frac{-i\gamma^\mu\, k_\mu}{(k^2)^{s/2}}\,.
\end{equation}
This gives\footnote{One can verify this expression using the Fourier transform
\begin{equation}
\int \frac{d^dk}{(2\pi)^d}\,e^{ik\cdot x}\,\frac{1}{(k^2)^{\frac{d}{2}-\Delta}}
=\frac{2^{2\Delta-d}}{\pi^\frac{d}{2}}\,\frac{\Gamma(\Delta)}
{\Gamma\left(\frac{d}{2}-\Delta\right)}\,\frac{1}{|x|^{2\Delta}}\,.
\end{equation}}
\begin{equation}
\label{Cf(s)}
C_f(s) = - \frac{2^{s-1}\,\Gamma\left(\frac{d+s}{2}\right)}{\pi^\frac{d}{2}\,
\Gamma\left(1-\frac{s}{2}\right)}\,.
\end{equation}
When $s=2$, we recover the usual (short-range) Dirac fermion propagator,
while the unitarity bound imposes the constraint
\begin{equation}
\label{unitarity for LR fermion}
s\leq 2\,.
\end{equation}
In position space, for general $s$, we obtain
\begin{equation}
\label{free gen fermion prop}
\langle \psi(x)\bar\psi(0)\rangle = C_\psi\,\frac{\gamma^\mu\, x_\mu}{|x|^{2\Delta_\psi+1}}\,,
\end{equation}
where $\Delta_\psi$ is defined by (\ref{definition of Delta_psi}), and the free propagator
amplitude is given by
\begin{equation}
\label{definition of C_psi}
C_\psi = \frac{\Gamma\left(\frac{d-s}{2}+1\right)}{2^{s-1}\,\pi^\frac{d}{2}\,\Gamma\left(\frac{s}{2}\right)}\,.
\end{equation}
Notice that when $s\rightarrow 2$ we recover the corresponding quantities
of the free short-range Dirac fermion\footnote{See \cite{Goykhman:2020ffn} for a recent review.}
\begin{equation}
\Delta_\psi|_{s\rightarrow 2} = \Delta_{\hat\psi}\,,\qquad
C_\psi|_{s\rightarrow 2} = C_{\hat\psi}\,.
\end{equation}

Let us now deform the action (\ref{free generalized psi action}) by a quartic
(Gross-Neveu) fermion self-interaction term,\footnote{
When $s=1$, the model can be seen as a free fermion in $d+1$-dimensional bulk, with the interaction localized
on the boundary. See also \cite{DiPietro:2020fya} for discussion of dualities between fixed points
of free vector models with interactions localized on the boundary.}
\begin{equation}
\label{def of LR GN}
S_{\textrm{LR\;GN}} = S_{\textrm{LR\;Dirac}} + \frac{g}{N}\,
\int d^dx \,(\bar\psi\psi)^2\,.
\end{equation}
In the vicinity of the free regime (\ref{free generalized psi action})
we obtain $[g] = 2s-2-d$. Therefore the quartic interaction in (\ref{def of LR GN})
is a relevant perturbation when $s> d/2+1$. Together with the unitarity constraint
(\ref{unitarity for LR fermion}), this imposes $d < 2$. In this case, the GN interaction
will drive the theory to a non-trivial regime in the IR. For $s=d/2+1+\epsilon$
the interaction strength is weak, and
one can perform a Wilson-Fisher perturbative $\epsilon$-expansion \cite{Wilson:1971dc}.

Instead, in this
paper we are going to focus on the domain where $s < d/2 + 1$, that results
in an irrelevant GN interaction term, driving the theory to an interacting regime in the UV.
Together with the unitarity constraint (\ref{unitarity for LR fermion}), we obtain
\begin{equation}
\label{upper bound on s prelim}
s < \min\left(\frac{d}{2}+1,2\right)\,.
\end{equation}
Therefore, when $d< 2$, we achieve a non-trivial regime in the UV for $s < d/2+1$
and in the IR for $s > d/2+1$. For $d>2$, we can only achieve a non-trivial regime
in the UV, for $s<2$.

Analogously to the bosonic case  \cite{Sak:1973,Sak:1977}, the long-range
model crosses over to the short-range regime for 
$s>s_\star$, where the threshold value of the exponent parameter is given by
\begin{equation}
s_\star = 2 - 2\gamma_{\hat\psi}\,.
\end{equation}
Due to $\gamma_{\hat\psi} > 0$, we obtain from (\ref{upper bound on s prelim})
\begin{equation}
\label{upper bound on s}
s < \min\left(\frac{d}{2}+1,s_\star\right)\,.
\end{equation}

To perform large-$N$ analysis of the long-range GN model (\ref{def of LR GN}),
it is convenient to introduce the Hubbard-Stratonovich pseudo-scalar field $\sigma\sim\bar\psi\psi$,
and rewrite the action as
\begin{equation}
\label{HS form of LR GN}
S_{\textrm{LR\;GN}} = S_{\textrm{LR\;Dirac}} +
\int d^dx \,\left(-\frac{1}{4g}\,\sigma^2+\frac{1}{\sqrt{N}}\,\sigma\,\bar\psi\psi\right)\,.
\end{equation}
Importantly, the model (\ref{HS form of LR GN}) enjoys the discrete $\mathbb{Z}_2$ symmetry,
\begin{equation}
\label{Z2 sym of LR GN}
\begin{aligned}
&(x^1,\dots,x^{a-1},x^a,x^{a+1},\dots,x^d)\,\rightarrow\, (x^1,\dots,x^{a-1},-x^a,x^{a+1},\dots,x^d)\,,\\
&\sigma\,\rightarrow\,-\sigma\,,\quad\psi\,\rightarrow\, \gamma^a\psi\,,\quad
\bar\psi\,\rightarrow\,-\bar\psi\gamma^a\,,
\end{aligned}
\end{equation}
defined for any given $a=1,\dots,d$. This is analogous to the local (short-range)
Gross-Neveu model (\ref{GN action HS}), that possesses the same symmetry
\cite{Gross:1974jv,Zinn-Justin:1991ksq,Moshe:2003xn}.

Integrating out the fermions, we arrive at the effective action for the Hubbard-Stratonovich field:
\begin{equation}
\label{LR GN with integrated out psi}
\begin{aligned}
S_{\textrm{LR\;GN}} &= -n\,\int d^dx\int d^dy\,\textrm{tr}\,\log
\left(C_f(s)\,\frac{(x-y)_\mu\gamma^\mu}{|x-y|^{d+s}}+\frac{1}{\sqrt{N}}\,\sigma(x)\,
\delta^{(d)}(x-y)\right)\\
&-\frac{1}{4g}\,\int d^dx \,\sigma^2\,,
\end{aligned}
\end{equation}
where trace is taken over gamma-matrices.
Expanding the logarithm in (\ref{LR GN with integrated out psi}) to leading order in $1/N$,
while discarding the constant and tadpole terms, we obtain\footnote{Here we use the free fermionic
inverse propagator relation:
\begin{equation}
\int d^dz\,\frac{C_f(s)\,(x-z)^\mu\gamma_\mu}{|x-z|^{d+s}}\,\frac{C_\psi\,z^\nu\gamma_\nu}{|z|^{d-s+2}}
=\delta^{(d)}(x)\,\mathbb{I}\,,
\end{equation}
that can be verified, for instance, using the Fourier transform.}
\begin{equation}
\label{LR GN effective action quadratic sigma}
S_{\textrm{LR\;GN}} = \frac{C_\psi^2}{2}\,\int d^dx\int d^dy\,\frac{\sigma(x)\sigma(y)}
{|x-y|^{2(d-s+1)}}-\frac{1}{4g}\,\int d^dx \,\sigma^2 + {\cal O}\left(\frac{1}{N}\right)\,.
\end{equation}
At the interacting fixed point in the UV, the first term in (\ref{LR GN effective action quadratic sigma})
dominates over the second term. Indeed, the corresponding propagator for the 
Hubbard-Stratonovich field (at the leading order in $1/N$) is given by
\begin{equation}
\label{leading <sigma sigma>}
\langle\sigma(x)\sigma(0)\rangle = \frac{C_\sigma}{|x|^{2\Delta_\sigma}}\,,
\end{equation}
where we defined the free propagator amplitude
\begin{equation}
C_\sigma =-\frac{4^{s-1} \Gamma (s-1) \Gamma \left(\frac{s}{2}\right)^2 \Gamma (d-s+1)}{\Gamma \left(\frac{d}{2}-s+1\right) \Gamma \left(\frac{d-s}{2} +1\right)^2 \Gamma \left(-\frac{d}{2}+s-1\right)}\,,
\end{equation}
and scaling dimension
\begin{equation}
\label{Delta sigma}
\Delta_\sigma = s-1\,.
\end{equation}
In other words, at the interacting UV f.p. we obtain
\begin{equation}
\label{[sigma] UV}
[\sigma] _ {\textrm{UV}} = s-1+{\cal O}\left(\frac{1}{N}\right)\,.
\end{equation}
We then obtain $[\sigma^2]_{\textrm{UV}} = 2(s-1)+{\cal O}\left(\frac{1}{N}\right) < d$. Therefore the 
second term in (\ref{LR GN effective action quadratic sigma}) is irrelevant in the UV.
As a consistency check, in the $s\rightarrow 2$ limit we can reproduce the corresponding quantities at the UV
fixed point of the local (short-range) GN model \cite{Zinn-Justin:1991ksq,Manashov:2017rrx,Manashov:2016uam},
\begin{equation}
\Delta_\sigma|_{s\rightarrow 2} = \Delta_{\hat\sigma}\,,\qquad
C_\sigma|_{s\rightarrow 2} = C_{\hat\sigma}\,.
\end{equation}

Demanding unitarity, we impose $[\sigma]_{\textrm{UV}} > \frac{d}{2}-1$, that in turn
demands $s > d/2$. Combing with (\ref{upper bound on s}), we obtain
\begin{equation}
\label{bounds on s}
\frac{d}{2}< s < \min\left(\frac{d}{2}+1,s_\star\right)\,,
\end{equation}
and therefore
\begin{equation}
\label{bounds on d}
1 < d < 4\,.
\end{equation}

Having established propagators for the fermion and the Hubbard-Stratonovich fields,
as well as the cubic interaction vertex, at 
the leading order in $1/N$, we return to the formulation of the model
in terms of the action (\ref{HS form of LR GN}).
The corresponding Feynman rules
that will be used at the interacting UV fixed point in diagrammatic calculations in this section are given by
\begin{center}
  \begin{picture}(452,62) (9,50)
    \SetWidth{1.0}
    \SetColor{Black}
    \Vertex(30,104){2}
    \Line[arrow,arrowpos=0.5,arrowlength=5,arrowwidth=2,arrowinset=0.2](30,104)(132,104)
    \Vertex(132,104){2}
    \Text(145,104)[lb]{\scalebox{0.8}{$x$}}
    \Text(172,94)[lb]{\scalebox{1}{$=C_\psi\,\frac{x^\mu\gamma_\mu}{|x|^{2\Delta_\psi+1}}$}}
    \Text(15,104)[lb]{\scalebox{0.8}{$0$}}
    \Vertex(30,64){2}
    \Line[](30,64)(132,64)
    \Vertex(132,64){2}
    \Text(145,64)[lb]{\scalebox{0.8}{$x$}}
    \Text(172,54)[lb]{\scalebox{1}{$=C_\sigma\,\frac{1}{|x|^{2\Delta_\sigma}}$}}
    \Text(15,64)[lb]{\scalebox{0.8}{$0$}}
    \Line[arrow,arrowpos=0.5,arrowlength=5,arrowwidth=2,arrowinset=0.2](348,84)(366,100)
    \Line[arrow,arrowpos=0.5,arrowlength=5,arrowwidth=2,arrowinset=0.2](366,66)(348,84)
    \Line[](348,84)(325,84)
    \Vertex(348,84){4}
    \Text(400,77)[lb]{\scalebox{1}{$=-\frac{1}{\sqrt{N}}$}}
  \end{picture}
\end{center}
When exponents of the propagator lines are specified explicitly,
the corresponding propagators will be implied to have unit amplitudes:
\begin{center}
  \begin{picture}(452,12) (9,95)
    \SetWidth{1.0}
    \SetColor{Black}
    \Vertex(30,104){2}
    \Line[arrow,arrowpos=0.5,arrowlength=5,arrowwidth=2,arrowinset=0.2](30,104)(132,104)
    \Vertex(132,104){2}
    \Text(145,104)[lb]{\scalebox{0.8}{$x$}}
    \Text(172,94)[lb]{\scalebox{1}{$=\frac{x^\mu\gamma_\mu}{|x|^{2\Delta+1}}$}}
    \Text(15,104)[lb]{\scalebox{0.8}{$0$}}
    \Text(80,114)[lb]{\scalebox{0.8}{$2\Delta$}}
    \Vertex(260,104){2}
    \Line[](260,104)(362,104)
    \Vertex(362,104){2}
    \Text(375,104)[lb]{\scalebox{0.8}{$x$}}
    \Text(402,94)[lb]{\scalebox{1}{$=\frac{1}{|x|^{2\Delta}}$}}
    \Text(245,104)[lb]{\scalebox{0.8}{$0$}}
    \Text(310,114)[lb]{\scalebox{0.8}{$2\Delta$}}
  \end{picture}
\end{center}
Recall also that each closed fermionic line produces the factor of $-1$ due to the anti-commuting nature of fermions.

\subsection{$\langle\psi\bar\psi\rangle$}
\label{sec: GN psi prop}

At the leading order in $1/N$ expansion,
the $\langle\psi\bar\psi\rangle$ propagator is given by
the generalized free field theory expression (\ref{free gen fermion prop}).
The scaling dimension $\Delta_\psi$, given by (\ref{definition of Delta_psi}),
is protected from loop corrections by the bi-local kinetic term, and therefore the anomalous dimension of $\psi$
vanishes, $\gamma_\psi = 0$, to all orders in $1/N$ expansion.
However, while  in the free limit the amplitude of the  propagator $C_\psi$ is given by (\ref{definition of C_psi}),
the $1/N$ corrections are going to renormalize it, resulting in
\begin{equation}
\langle \psi(x)\bar\psi(0)\rangle = C_\psi\,(1+A_\psi)\,\frac{\gamma^\mu\, x_\mu}{|x|^{2\Delta_\psi+1}}\,.
\end{equation}
Our goal in this section is to find the relative correction
to the propagator amplitude $A_\psi$ at the next-to-leading order in $1/N$
expansion. 
The only contributing diagram is one-loop, given by:
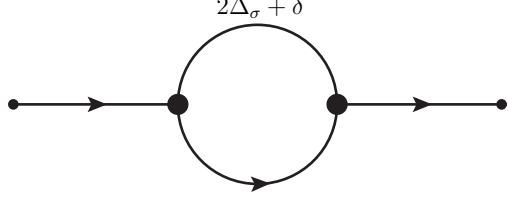
\begin{center}
  \begin{picture}(296,70) (-20,10)
    \SetWidth{1.0}
    \SetColor{Black}
    \Arc[arrow,arrowpos=0.5,arrowlength=5,arrowwidth=2,arrowinset=0.2](130,44)(30,180,364)
    \Arc[](130,44)(30,0,180)
    \Line[arrow,arrowpos=0.5,arrowlength=5,arrowwidth=2,arrowinset=0.2](38,44)(101,44)
    \Line[arrow,arrowpos=0.5,arrowlength=5,arrowwidth=2,arrowinset=0.2](160,44)(224,44)
    \Vertex(100,44){4.001}
    \Vertex(160,44){4.001}
    \Vertex(38,44){2.001}
    \Vertex(222,44){2.001}
    \Text(115,77)[lb]{\scalebox{0.8}{$2\Delta_\sigma+\delta$}}
  \end{picture}
\end{center}
The standard technique to renormalize such a conformal graph is to add a small
shift $\delta$ to the exponents of the internal $\sigma$ lines \cite{Vasiliev:1975mq}.
Subsequently using the loop identities in position space,
as well as the propagator merging relations,
reviewed in Appendix~\ref{app_a}, we obtain
\begin{equation}
\begin{aligned}
\langle \psi(x)\bar\psi(0)\rangle&\supset
-\frac{1}{N}C_\psi^3C_\sigma \pi^d A\left(-\frac{\delta}{2}\right)
V\left(\frac{d+s-1+\delta}{2},\frac{d-s+1}{2}\right)\\
&\times A\left(\frac{d+\delta}{2}\right)
V\left(\frac{d-s+1}{2},\frac{s-1-\delta}{2}\right)\frac{x^\mu\gamma_\mu}
{|x|^{d-s+2+\delta}}\,.
\end{aligned}
\end{equation}
This expression is in fact finite in the $\delta\rightarrow 0$ limit, in agreement with
our expectation that $\gamma_\psi = 0$. Taking this limit renders
\begin{equation}
\label{A psi}
A_\psi = \frac{1}{N}\frac{2^{d-s-1} (d-2 s+2)}{\sqrt{\pi } \Gamma \left(\frac{d+s}{2}\right)
\sin \left(\frac{\pi  s}{2}\right)}
  \sin \left(\frac{ \pi  (d-2 s)}{2}\right) \Gamma (s-1) \Gamma \left(\frac{d-s+1}{2} \right)\,.
\end{equation}
While propagator amplitudes are not a part of universal CFT data,
in combination with the other non-observable, given by the amplitude
of the $\sigma\bar\psi\psi$ conformal triangle, discussed in section~\ref{sec: GN sigma psi psi conformal triangle},
the expression (\ref{A psi}) will play an important role in calculation of the OPE
coefficient $\langle \sigma\bar\psi\psi\rangle$, that we are going to carry out in
section~\ref{sec: GN sigma psi psi}.

\subsection{$\langle\sigma\sigma\rangle$}
\label{sec: GN sigma prop}

At the interacting UV fixed point, dimension of the Hubbard-Stratonovich
field, $\sigma\simeq \bar\psi\psi$, is given by (\ref{[sigma] UV}),
where the leading order in $1/N$ contribution is given by (\ref{Delta sigma}).
When $1/N$ corrections are taken into account, the 
leading order $\langle\sigma\sigma\rangle$ propagator (\ref{leading <sigma sigma>})
gets modified, acquiring the form
\begin{equation}
\langle\sigma (x)\sigma (0)\rangle
=\frac{C_\sigma\,(1+A_\sigma)\mu^{-2\gamma_\sigma}}{|x|^{2(\Delta_\sigma+\gamma_\sigma)}}\,,
\end{equation}
where $\mu$ is an arbitrary scale.
Our goal in this section is to find anomalous dimension $\gamma_\sigma$
at the next-to-leading order in $1/N$ expansion. This constitutes 
an important part of CFT data of the corresponding critical theory.
We will also calculate the relative propagator amplitude correction
$A_\sigma$, that will play a role in calculation of
the  $\langle \sigma\bar\psi\psi\rangle$  three-point
function in section~\ref{sec: GN sigma psi psi}.

The two-loop diagrams contributing to the $\langle\sigma\sigma\rangle$
propagator at the next-to-leading order in $1/N$ expansion are given by:\footnote{One
more $\langle\sigma\sigma\rangle$ two-loop diagram at the order ${\cal O}(1/N)$
can be drawn, but it vanishes, since it contains
the $\langle\sigma\sigma\sigma\rangle$ triangle sub-diagram. The latter correlation
function is in fact identically zero in the Gross-Neveu model,
both in the short-range \cite{Manashov:2016uam,Goykhman:2020ffn}, and in the long-range case,
due to the $\mathbb{Z}_2$ symmetry
(\ref{Z2 sym of LR GN}). This makes some calculations in the Gross-Neveu
model easier than in the bosonic $O(N)$
vector model. In the latter, vanishing of the $\langle\sigma\sigma\sigma\rangle$
three-point function is a three-dimensional artifact \cite{Petkou:1995vu,Petkou:1994ad,Goykhman:2019kcj}, and the
counterpart of the $\mathbb{Z}_2$ symmetry does not seem to generally hold 
beyond the leading order in $1/N$ expansion \cite{Chai:2021uhv}.
}
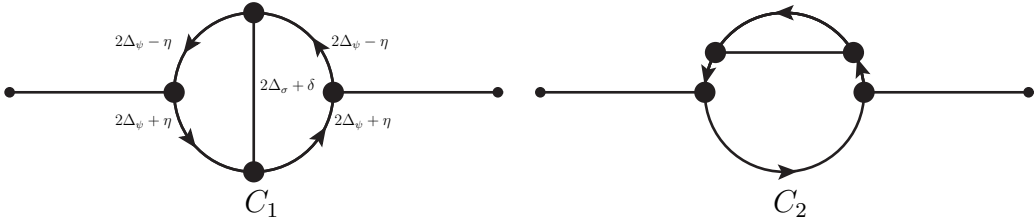
\begin{center}
  \begin{picture}(296,70) (90,10)
    \SetWidth{1.0}
    \SetColor{Black}
    \Arc[arrow,arrowpos=0.2,arrowlength=5,arrowwidth=2,arrowinset=0.2](130,44)(30,180,364)
    \Arc[arrow,arrowpos=0.8,arrowlength=5,arrowwidth=2,arrowinset=0.2](130,44)(30,180,364)
    \Arc[arrow,arrowpos=0.2,arrowlength=5,arrowwidth=2,arrowinset=0.2](130,44)(30,0,180)
    \Arc[arrow,arrowpos=0.8,arrowlength=5,arrowwidth=2,arrowinset=0.2](130,44)(30,0,180)
    \Line[](38,44)(101,44)
    \Line[](160,44)(224,44)
    \Line[](130,14)(130,74)
    \Vertex(130,14){4.001}
    \Vertex(130,74){4.001}
    \Vertex(100,44){4}
    \Vertex(160,44){4}
    \Vertex(38,44){2.001}
    \Vertex(222,44){2}
    \Text(133,44)[lb]{\scalebox{0.5}{$2\Delta_\sigma+\delta$}}
    \Text(78,60)[lb]{\scalebox{0.5}{$2\Delta_\psi-\eta$}}
    \Text(160,60)[lb]{\scalebox{0.5}{$2\Delta_\psi-\eta$}}
    \Text(78,30)[lb]{\scalebox{0.5}{$2\Delta_\psi+\eta$}}
    \Text(161,30)[lb]{\scalebox{0.5}{$2\Delta_\psi+\eta$}}
    \Text(127,-3)[lb]{\scalebox{1}{$C_1$}}
    \Arc[arrow,arrowpos=0.5,arrowlength=5,arrowwidth=2,arrowinset=0.2](330,44)(30,180,364)
    \Arc[arrow,arrowpos=0.1,arrowlength=5,arrowwidth=2,arrowinset=0.2](330,44)(30,0,180)
    \Arc[arrow,arrowpos=0.5,arrowlength=5,arrowwidth=2,arrowinset=0.2](330,44)(30,0,180)
    \Arc[arrow,arrowpos=0.93,arrowlength=5,arrowwidth=2,arrowinset=0.2](330,44)(30,0,180)
    \Line[](238,44)(301,44)
    \Line[](360,44)(424,44)
    \Vertex(304,59){4.001}
    \Vertex(356,59){4}
    \Line[](304,59)(356,59)
    \Vertex(300,44){4.001}
    \Vertex(360,44){4}
    \Vertex(238,44){2}
    \Vertex(422,44){2}
     \Text(327,-3)[lb]{\scalebox{1}{$C_2$}}
    %
  \end{picture}
\end{center}
We are going to denote the corresponding expressions
contributing to $\langle\sigma\sigma\rangle$ as $C_{1,2}$.
The diagram $C_1$ 
is divergent, and requires regularization that we accomplish
by adding a small shift $\delta$ to the exponent of the
internal $\sigma$ line \cite{Vasiliev:1975mq}. Presence of $\delta$
spoils uniqueness of the three-vertices. However, using an auxiliary
regulator $\eta = \delta/2$ one can restore the integrability of the diagram,
without affecting its regularized value. This is due
to the fact that the diagram possesses the $\eta\rightarrow-\eta$ symmetry,\footnote{
This can be seen by renaming positions of vertices of integration,
$x_{3,4}\rightarrow x_1+x_2-x_{4,3}$, where $x_{1,2}$ are left and right vertices,
and $x_{3,4}$ are top and bottom vertices.}
and therefore $\eta$ dependence can only appear as $1+{\cal O}(\eta^2)$.
Since the strongest divergence of this diagram is the simple pole $1/\delta$,
any choice $\eta={\cal O}(\delta)$ will not influence the $\delta\rightarrow 0$ limit \cite{Vasiliev:1981yc,Vasiliev:1981dg,Belokurov:1983rkp,Ciuchini:1999wy,Gubser:2017vgc}.

Using uniqueness and propagator merging relations, reviewed in
Appendix~\ref{app_a}, we obtain\footnote{Here the overall factor of $-1$
is due to the Feynman rule for the fermionic loop.}
\begin{align}
C_1 &=-\frac{1}{N} C_\psi^4C_\sigma^3
\pi^\frac{d}{2}A\left(s-1+\frac{\delta}{2}\right)
V\left(\frac{d-s+1}{2}-\frac{\delta}{4},\frac{d-s+1}{2}-\frac{\delta}{4}\right)
U\left(\frac{d+\delta}{2},\frac{d+\delta}{2},-\delta\right)\notag\\
&\times U\left(s-1,d-s+1+\frac{\delta}{2},-\frac{\delta}{2}\right)
U\left(\frac{d+\delta}{2},s-1,\frac{d-\delta}{2}-s+1\right)\frac{\mu^{-\delta}}{|x|^{2s-2+\delta}}\,.
\end{align}
Expanding around $\delta = 0$, we obtain\footnote{Simple pole divergence
can be removed by the wave-function renormalization
$\sigma\rightarrow\sqrt{1+\frac{2\gamma_\sigma}{\delta}}\sigma$.
For brevity, and where it does not create confusion, we keep subtractions of
pure infinities implicit in this paper.}
\begin{equation}
\label{gamma sigma result}
\gamma_\sigma = \frac{1}{N}
\frac{2 \Gamma \left(\frac{s}{2}\right)^2 \Gamma (d-s+1)}
{\Gamma \left(\frac{d}{2}\right) \Gamma \left(\frac{d-s+2}{2} \right)^2 \Gamma \left(-\frac{d}{2}+s-1\right)}
+{\cal O}\left(\frac{1}{N^2}\right)\,.
\end{equation}
The finite part is given by
\begin{align}
\label{C1}
\frac{C_1}{C_\sigma}\supset \frac{1}{N}\frac{2 \Gamma \left(\frac{s}{2}\right)^2 \Gamma (d{-}s{+}1) \left(H_{\frac{d{-}s}{2}}{-}H_{d{-}s}{-}\psi ^{(0)}\left({-}\frac{d}{2}{+}s{-}1\right){+}\psi ^{(0)}\left(\frac{s}{2}\right)\right)}{ \Gamma \left(\frac{d}{2}\right) \Gamma \left(\frac{d{-}s}{2} {+}1\right)^2 \Gamma \left({-}\frac{d}{2}{+}s{-}1\right)}\equiv c_1\,,
\end{align}
where $\psi^{(n)}(x)$ is $n$th derivative of the digamma function $\psi^{(0)}(x)=\Gamma'(x)/\Gamma(x)$.

The diagram $C_2$ contains the finite one-loop correction to 
the fermionic propagator $\langle\psi\bar\psi\rangle$ as a sub-diagram:\footnote{
Here we took into account the symmetry factor of $2$,
while the factor of $-1$ due to the fermionic loop was offset
by the factor of $-1$ due to the simplification
of loop in position space, reviewed in Appendix~\ref{app_a}.
We have then used the inverse propagator relation
\begin{equation}
\int d^dy\frac{1}{|y|^{2a}|x-y|^{2(d-a)}}=\pi^d A(a)A(d-a)\,.
\end{equation}
}
\begin{equation}
\label{C2}
\begin{aligned}
\frac{C_2}{C_\sigma} &\supset 2\,C_\psi^2C_\sigma\, A_\psi \,\pi^d\, A(s-1)A(d-s+1)\\
&=-\frac{1}{N}\frac{2^{d-s} (d-2 s+2)\Gamma (s-1) \sin \left(\frac{1}{2} \pi  (d-2 s)\right) \Gamma \left(\frac{1}{2} (d-s+1)\right)}{\sqrt{\pi }  \sin \left(\frac{\pi  s}{2}\right)  \Gamma \left(\frac{d+s}{2}\right)}\equiv c_2\,,
\end{aligned}
\end{equation}
and therefore does not contribute to the anomalous dimension $\gamma_\sigma$.
Combining (\ref{C1}), (\ref{C2}), we obtain
\begin{equation}
\label{A sigma}
A_\sigma = c_1 + c_2\,.
\end{equation}

We close this section by providing an expansion of the anomalous dimension $\gamma_\sigma$,
given by (\ref{gamma sigma result}), at $d=4-\epsilon_2$, $s=2-\epsilon_2/2+\epsilon_1$
to leading order in $\epsilon_{1,2}$,
\begin{equation}
\label{gamma sigma expanded result}
\gamma_\sigma|_{d=4} = - \frac{4}{N}\,\epsilon_1+{\cal O}\left(\epsilon_{1,2}^2,\frac{1}{N^2}\right)\,.
\end{equation}
 This expansion will prove useful when we compare the CFT data of the long-range Gross-Neveu
 model with its counterpart in the Gross-Neveu-Yukawa model.

\subsection{$\sigma\bar\psi\psi$ conformal triangle}
\label{sec: GN sigma psi psi conformal triangle}

While the leading order $\sigma\bar\psi\psi$ vertex in the action (\ref{HS form of LR GN}) is local,
higher-order $1/N$ corrections result in a non-local effective vertex for the cubic  $\sigma\bar\psi\psi$
interaction. Diagrammatically, in a CFT, it
can be represented using the conformal triangle \cite{Polyakov:1970xd}
\begin{center}
\begin{equation}
\label{definition of vertex Gamma}
  \begin{picture}(300,97) (73,3)
    \SetWidth{1.0}
    \SetColor{Black}
    \Line[arrow,arrowpos=0.5,arrowlength=5,arrowwidth=2,arrowinset=0.2](80,95)(52,47)
    \Line[](52,47)(108,47)
    \Line[arrow,arrowpos=0.5,arrowlength=5,arrowwidth=2,arrowinset=0.2](111,44)(83,96)
    \Vertex(81,97){4}
    \Vertex(52,47){4}
    \Vertex(111,47){4}
    \Text(69,107)[lb]{\scalebox{1}{$\sigma(x_3)$}}
    \Text(20,35)[lb]{\scalebox{1}{$\bar\psi(x_1)$}}
    \Text(115,35)[lb]{\scalebox{01}{$\psi(x_2)$}}
    \Text(102,73)[lb]{\scalebox{0.8001}{$2\alpha$}}
    \Text(78,34)[lb]{\scalebox{0.8001}{$2\beta$}}
    \Text(50,73)[lb]{\scalebox{0.8001}{$2\alpha$}}
    \Text(120,65)[lb]{\scalebox{0.9001}{$~=~-\frac{Z_{\sigma\bar\psi\psi}}{\sqrt{N}}\,\int d^dx_1\int d^dx_2\int d^dx_3\,
\frac{\mu^{ \gamma_\sigma}\;x_{13}^\mu\gamma_\mu \; x_{32}^\nu\gamma_\nu}
{|x_{13}|^{2\alpha+1}|x_{32}|^{2\alpha+1}|x_{12}|^{2\beta}}\,\bar\psi(x_1)\psi(x_2)\sigma(x_3)\,,$}}
  \end{picture}
  \end{equation}
\end{center}
Exponents $2\alpha$, $2\beta$ of  internal lines of the $\sigma\bar\psi\psi$ conformal triangle
have been chosen so that when the full $\sigma$ and $\psi$ propagators are attached to it,
each of the three internal vertices of the triangle are rendered unique:
\begin{equation}
\alpha = \frac{d-s+1-\gamma_\sigma}{2}\,,\qquad \beta = s-1+\frac{\gamma_\sigma}{2}\,.
\end{equation}
Performing the resulting three unique integrals then generates the factor of
\begin{equation}
\begin{aligned}
&\textrm{integrals of }(\ref{definition of vertex Gamma})=-\pi^\frac{3d}{2}A(s{-}1{+}\gamma_\sigma)
V\left(\frac{d{-}s{+}1{-}\gamma_\sigma}{2},\frac{d{-}s{+}1{-}\gamma_\sigma}{2}\right)\\
&\times  A\left(\frac{d{-}\gamma_\sigma}{2}\right)V\left(\frac{d{-}s{+}1}{2},\frac{s{-}1{+}\gamma_\sigma}{2}\right)
A\left(s{-}1{+}\frac{\gamma_\sigma}{2}\right)
V\left(\frac{d{-}s{+}1}{2},\frac{d{-}s{+}1{-}\gamma_\sigma}{2}\right)\,.
\end{aligned}
\end{equation}
Expanding in $1/N$, we can re-write this factor as
\begin{equation}
\label{integral over vertices of conf triangle}
\textrm{integrals of }(\ref{definition of vertex Gamma})=u_{\sigma\bar\psi\psi}^{(0)}\left(1+\delta u_{\sigma\bar\psi\psi}
+{\cal O}\left(\frac{1}{N^2}\right)\right)\,,
\end{equation}
where we denoted
\begin{align}
u_{\sigma\bar\psi\psi}^{(0)}&{=}-N\,\frac{\pi ^{\frac{3 d}{2}} \Gamma \left(\frac{s}{2}\right)^2 \Gamma \left(\frac{d}{2}{-}s{+}1\right)^2 \Gamma \left({-}\frac{d}{2}{+}s{-}1\right)}{\Gamma (s{-}1)^2 \Gamma (d{-}s{+}1) \Gamma \left(\frac{d{-}s}{2} {+}1\right)^2}\,,\\
\delta u_{\sigma\bar\psi\psi}&{=}\frac{1}{N}\frac{\Gamma \left(\frac{s}{2}\right)^2 \Gamma (d{-}s{+}1) \left(2 H_{\frac{d-s}{2}}{-}3 H_{\frac{d}{2}{-}s}+\psi ^{(0)}\left(\frac{d}{2}\right){+}2 \psi ^{(0)}\left(\frac{s}{2}\right){-}3 \psi ^{(0)}(s{-}1)\right)}{ \Gamma \left(\frac{d}{2}\right) \Gamma \left(\frac{d{-}s}{2} {+}1\right)^2 \Gamma \left({-}\frac{d}{2}{+}s{-}1\right)}\,.
\end{align}
Additionally taking into account the propagator amplitude corrections\footnote{
Importance of the propagator amplitude corrections in calculation of three-point functions
in large-$N$ vector models was first emphasized in \cite{Goykhman:2019kcj}.}
one can obtain the three-point
function $\langle\sigma\bar\psi\psi\rangle$. This calculation will be completed in section~\ref{sec: GN sigma psi psi}.

Conformal triangle can be calculated at the desired order in $1/N$ expansion
by summing up all contributing diagrams up to the given order. These diagrams include vertex corrections,
propagator corrections, and counterterms. Many of these diagrams contain dressed or regularized
propagators, rendering them non-integrable via the method of uniqueness. In \cite{Goykhman:2020ffn}
the background field method was proposed,
that allows to easily calculate conformal triangles by using the propagator merging relations.\footnote{The
background field method was recently used to calculate CFT data in the long-range $O(N)$
vector model \cite{Chai:2021arp}.}
In such an approach, one of the fields is formally fixed to a non-dynamical background value.
For instance, one can fix the Hubbard-Stratonovich
field to a constant $\sigma \equiv \bar\sigma$, while attaching the full fermionic propagators
to the $\sigma\bar\psi\psi$ conformal triangle. The conformal triangle can then be solved for,
by considering two equivalent forms of writing down the fermionic propagator $\langle\psi\bar\psi\rangle|_{\bar\sigma}$ in the
Hubbard-Stratonovich background $\bar\sigma$:
\begin{center}
\begin{equation}
  \begin{picture}(500,50) (6,120)
    \SetWidth{1.0}
    \SetColor{Black}
    \Text(82,177)[lb]{\scalebox{1.001}{$\bar \sigma$}}
    \Text(62,157)[lb]{\scalebox{0.6}{$2\alpha$}}
    \Text(100,157)[lb]{\scalebox{0.6}{$2\alpha$}}
    \Text(82,132)[lb]{\scalebox{0.6001}{$2\beta$}}
    \Vertex(85,167){4}
    \Line[arrow,arrowpos=0.5,arrowlength=5,arrowwidth=2,arrowinset=0.2](85,167)(65,140)
    \Vertex(65,140){4}
    \Line[arrow,arrowpos=0.5,arrowlength=5,arrowwidth=2,arrowinset=0.2](104,141)(85,167)
    \Vertex(104,141){4}
    \Line[](66,141)(104,141)
    \Line[arrow,arrowpos=0.5,arrowlength=5,arrowwidth=2,arrowinset=0.2](65,140)(38,122)
    \Line[arrow,arrowpos=0.5,arrowlength=5,arrowwidth=2,arrowinset=0.2](130,122)(104,141)
    \Vertex(38,122){2}
    \Vertex(130,122){2}
    \Text(145,147)[lb]{\scalebox{1.001}{$=$}}
    \Text(195,168)[lb]{\scalebox{1}{$\bar \sigma$}}
    \Line[arrow,arrowpos=0.5,arrowlength=5,arrowwidth=2,arrowinset=0.2](197,159)(161,122)
    \Line[arrow,arrowpos=0.5,arrowlength=5,arrowwidth=2,arrowinset=0.2](230,122)(197,159)
    \Vertex(197,159){4}
    \Vertex(161,122){2}
    \Vertex(230,122){2}
    \Text(285,168)[lb]{\scalebox{1.001}{$\bar \sigma$}}
    \Text(230,142)[lb]{\scalebox{1}{$+\frac{\gamma_\sigma}{\delta}~\times$}}
    \Line[arrow,arrowpos=0.5,arrowlength=5,arrowwidth=2,arrowinset=0.2](287,159)(252,122)
    \Line[arrow,arrowpos=0.5,arrowlength=5,arrowwidth=2,arrowinset=0.2](321,122)(287,159)
    \Vertex(252,122){2}
    \Vertex(321,122){2}
    \Vertex(287,159){4}
    \Text(390,178)[lb]{\scalebox{1}{$\bar \sigma$}}
    \Text(331,145.5)[lb]{\scalebox{1}{$+$}}
    \Text(381,131)[lb]{\scalebox{0.6}{$2\Delta_\sigma+\delta$}}
    \Vertex(392,169){4}
    \Line[arrow,arrowpos=0.5,arrowlength=5,arrowwidth=2,arrowinset=0.2](392,169)(373,141)
    \Line[arrow,arrowpos=0.5,arrowlength=5,arrowwidth=2,arrowinset=0.2](412,141)(392,169)
    \Line[](373,141)(412,141)
    \Vertex(373,141){4}
    \Vertex(412,141){4}
    \Line[arrow,arrowpos=0.5,arrowlength=5,arrowwidth=2,arrowinset=0.2](373,141)(345,122)
    \Line[arrow,arrowpos=0.5,arrowlength=5,arrowwidth=2,arrowinset=0.2](438,122)(412,141)
    \Vertex(345,122){2}
    \Vertex(438,122){2}
  \end{picture}
   \label{Dressed fermion propagator in s background}
\end{equation}
\end{center}
On the l.h.s. of (\ref{Dressed fermion propagator in s background}) we obtain\footnote{It is convenient
to factor out $\bar\sigma C_\psi^2(1{+}A_\psi)^2$ from each term on both sides of (\ref{Dressed fermion propagator in s background}). Integrals over vertices
of the diagram on the l.h.s. of (\ref{Dressed fermion propagator in s background}) produces the factor equal to the negative of (\ref{integral over vertices of conf triangle}).}
\begin{align}
\label{simplified lhs of conf triangle}
\frac{\textrm{l.h.s.\;of\;} (\ref{Dressed fermion propagator in s background})}{\bar\sigma C_\psi^2(1+A_\psi)^2}
= \frac{Z_{\sigma\bar\psi\psi}^{(0)}}{\sqrt{N}}\,u_{\sigma\bar\psi\psi}^{(0)}\,
\left(1+\delta Z_{\sigma\bar\psi\psi}+\delta u_{\sigma\bar\psi\psi}+\gamma_\sigma\log(\mu |x|)\right)\,,
\end{align}
where we have also expanded amplitude of the conformal triangle in $1/N$,
\begin{equation}
Z_{\sigma\bar\psi\psi} = Z_{\sigma\bar\psi\psi}^{(0)}(1+\delta Z_{\sigma\bar\psi\psi})\,,
\end{equation}

For the first term on the r.h.s. of (\ref{Dressed fermion propagator in s background}) we derive
\begin{equation}
\label{tree level in conf triangle}
\frac{\langle\psi\bar\psi\rangle|_{\bar\sigma}}{\bar\sigma C_\psi^2(1{+}A_\psi)^2}\supset
-\frac{C_{\sigma\bar\psi\psi}^{(0)}}{C_\sigma^\frac{1}{2} C_\psi}\,
\frac{1}{|x|^{d-2s+2}}\,,
\end{equation}
where for the future convenience we introduced
\begin{equation}
\label{defintion of C sigma psi psi 0}
C_{\sigma\bar\psi\psi}^{(0)} = -\frac{1}{\sqrt{N}}\,C_\sigma^\frac{1}{2} C_\psi \,
\pi^\frac{d}{2}\,A(s-1)V\left(\frac{d-s+1}{2},\frac{d-s+1}{2}\right)\,.
\end{equation}
Comparing (\ref{tree level in conf triangle}) with the leading-order contribution to (\ref{simplified lhs of conf triangle}),
we obtain
\begin{equation}
\label{leading order Z0}
Z_{\sigma\bar\psi\psi}^{(0)}=-\sqrt{N}\frac{C_{\sigma\bar\psi\psi}^{(0)}}{C_\sigma^\frac{1}{2}C_\psi u_{\sigma\bar\psi\psi}^{(0)}}
=-\frac{1}{N}
\frac{\Gamma (s-1) \Gamma (d-s+1)}{\pi ^{d}  \Gamma \left(\frac{d}{2}-s+1\right) \Gamma \left(-\frac{d}{2}+s-1\right)}\,.
\end{equation}
At the same time, the sub-leading order contributions to (\ref{simplified lhs of conf triangle})
are to be compared to the
regularized vertex correction diagram, represented by the last term
in the r.h.s. of (\ref{Dressed fermion propagator in s background}), given by
\begin{align}
&\frac{\langle\psi\bar\psi\rangle|_{\bar\sigma}}{\bar\sigma C_\psi^2(1{+}A_\psi)^2}\supset
-\frac{1}{N^\frac{3}{2}}\,\pi^\frac{3d}{2}C_\psi^2C_\sigma A(s-1)V\left(\frac{d-s+1}{2},\frac{d-s+1}{2}\right)
A\left(\frac{d+\delta}{2}\right)\\
&\times V\left(\frac{d{-}s{+}1}{2},\frac{s{-}1{-}\delta}{2}\right)
A\left(s{-}1{-}\frac{\delta}{2}\right)V\left(\frac{d{-}s{+}1}{2},\frac{d{-}s{+}1{+}\delta}{2}\right)
\frac{\mu^{-\delta}}{|x|^{d{-}2s{+}2{+}\delta}}\,.
\end{align}
Expanding around $\delta = 0$, we notice that the pure pole divergence $1/\delta$
is subtracted by the counterterm diagram in the r.h.s. of (\ref{Dressed fermion propagator in s background}),
while the finite part assumes the form
\begin{align}
\label{loop correction to sigma psi psi vertex}
\frac{\langle\psi\bar\psi\rangle|_{\bar\sigma}}{\bar\sigma C_\psi^2(1{+}A_\psi)^2}\supset
-\frac{C_{\sigma\bar\psi\psi}^{(0)}}{C_\sigma^\frac{1}{2}C_\psi}\,\left(
r_{\sigma\bar\psi\psi}+\gamma_\sigma\,\log(\mu |x|)\right)\,,
\end{align}
where we denoted
\begin{equation}
r_{\sigma\bar\psi\psi}=\frac{1}{N}\,
\frac{\Gamma \left(\frac{s}{2}\right)^2 \Gamma (d-s+1) \left(\psi ^{(0)}\left(\frac{d}{2}\right)-H_{\frac{d}{2}-s}-H_{s-2}+\gamma \right)}{ \Gamma \left(\frac{d}{2}\right) \Gamma \left(\frac{d-s}{2} +1\right)^2 \Gamma \left(-\frac{d}{2}+s-1\right)}\,,
\end{equation}
where $\gamma$ is the Euler constant.
Comparing (\ref{loop correction to sigma psi psi vertex}) with the sub-leading contribution to
(\ref{simplified lhs of conf triangle}), while using (\ref{leading order Z0}), we notice that the anomalous
dimension logarithm terms match, while the relative correction to amplitude of the conformal
triangle can be expressed as
\begin{equation}
\begin{aligned}
\delta Z_{\sigma\bar\psi\psi} &= r_{\sigma\bar\psi\psi} - \delta u_{\sigma\bar\psi\psi}\\
&=\frac{1}{N}
\frac{2 \Gamma \left(\frac{s}{2}\right)^2 \Gamma (d-s+1) \left(H_{\frac{d}{2}-s}-H_{\frac{d-s}{2}}-\psi ^{(0)}\left(\frac{s}{2}\right)+\psi ^{(0)}(s-1)\right)}{\Gamma \left(\frac{d}{2}\right) \Gamma \left(\frac{d-s}{2} +1\right)^2 \Gamma \left(-\frac{d}{2}+s-1\right)}\,.
\end{aligned}
\end{equation}

\subsection{$\langle\sigma\bar\psi\psi\rangle$}
\label{sec: GN sigma psi psi}

At the leading order in $1/N$ expansion, the three-point function $\langle\sigma\bar\psi\psi\rangle$
is determined by the tree-level diagram
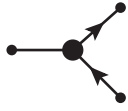
\begin{center}
  \begin{picture}(652,22) (140,67)
    \SetWidth{1.0}
    \SetColor{Black}
    \Vertex(366,100){2}
    \Vertex(366,66){2}
    \Vertex(325,84){2}
    \Line[arrow,arrowpos=0.5,arrowlength=5,arrowwidth=2,arrowinset=0.2](348,84)(366,100)
    \Line[arrow,arrowpos=0.5,arrowlength=5,arrowwidth=2,arrowinset=0.2](366,66)(348,84)
    \Line[](348,84)(325,84)
    \Vertex(348,84){4}
  \end{picture}
\end{center}
This diagram is easy to evaluate taking one conformal integral using
the uniqueness relation. This gives
\begin{equation}
\langle \sigma (x_1)\bar\psi(x_2)\psi (x_3)\rangle\Bigg|_{\textrm{normalized}}
=\frac{C_{\sigma\bar\psi\psi}^{(0)}\,x_{12}^\mu\gamma_\mu\,x_{31}^\nu\gamma_\nu}
{(|x_{12}||x_{13}|)^{s}|x_{23}|^{d-2s+2}}\,,
\end{equation}
where the amplitude $C_{\sigma\bar\psi\psi}^{(0)}$ is given by (\ref{defintion of C sigma psi psi 0}).
Following the same conventions throughout this paper, all three-point
functions are written down for operators whose propagators
in position space have been normalized to unity.

To incorporate $1/N$ contributions due to the vertex and propagator corrections, we can
use $\sigma\bar\psi\psi$ conformal triangle, calculated in section~\ref{sec: GN sigma psi psi},
and relative propagator amplitude corrections, calculated in section~\ref{sec: GN psi prop}
and section~\ref{sec: GN sigma prop}.
This gives the full three-point function
\begin{equation}
\langle \sigma (x_1)\bar\psi(x_2)\psi (x_3)\rangle\Bigg|_{\textrm{normalized}}
=\frac{C_{\sigma\bar\psi\psi}^{(0)}(1+\delta C_{\sigma\bar\psi\psi})\,x_{12}^\mu\gamma_\mu\,x_{31}^\nu\gamma_\nu}
{(|x_{12}||x_{13}|)^{s+\gamma_\sigma}|x_{23}|^{d-2s+2-\gamma_\sigma}}\,,
\end{equation}
Attaching full propagator legs to the conformal
triangle and taking integrals over the unique vertices, we obtain:
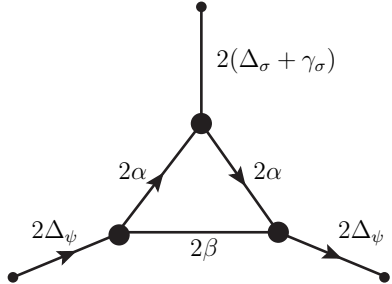
\begin{center}
  \begin{picture}(163,114) (0,-10)
    \SetWidth{1.0}
    \SetColor{Black}
    \Line[](81,95)(81,51)
    \Line[](50,10)(110,10)
    \Line[arrow,arrowpos=0.5,arrowlength=5,arrowwidth=2,arrowinset=0.2](10,-7)(51,10)
    \Line[arrow,arrowpos=0.5,arrowlength=5,arrowwidth=2,arrowinset=0.2](110,10)(151,-7)
    \Line[arrow,arrowpos=0.5,arrowlength=5,arrowwidth=2,arrowinset=0.2](48,8)(81,51)
    \Line[arrow,arrowpos=0.5,arrowlength=5,arrowwidth=2,arrowinset=0.2](81,51)(110,10)
    \Vertex(50,9){4.001}
    \Vertex(81,51){4}
    \Vertex(110,10){4}
    \Vertex(10,-7){2}
    \Vertex(150,-7){2}
    \Vertex(81,95){2.001}
    \Text(87,70)[lb]{\scalebox{0.8}{$2(\Delta_\sigma+\gamma_\sigma)$}}
    \Text(50,30)[lb]{\scalebox{0.8}{$2\alpha$}}
    \Text(101,30)[lb]{\scalebox{0.8}{$2\alpha$}}
    \Text(77,-1)[lb]{\scalebox{0.8}{$2\beta$}}
    \Text(17,5)[lb]{\scalebox{0.8}{$2\Delta_\psi$}}
    \Text(133,5)[lb]{\scalebox{0.8}{$2\Delta_\psi$}}
  \end{picture}
\end{center}
Therefore the relative  correction to the three-point function amplitude is given by
\begin{align}
\notag
\delta C_{\sigma\bar\psi\psi}&=\delta Z_{\sigma\bar\psi\psi}+\delta u_{\sigma\bar\psi\psi}+\frac{A_\sigma}{2}+A_\psi\\
&=-\frac{1}{N}\frac{\Gamma \left(\frac{s}{2}\right)^2 \Gamma (d-s+1) }
{ \Gamma \left(\frac{d}{2}\right) \Gamma \left(\frac{d-s}{2} +1\right)^2 \Gamma \left(-\frac{d}{2}+s-1\right)}\,
\left(H_{\frac{d}{2}-s}-H_{\frac{d-s}{2}}+H_{d-s}\right.\label{delta C sigma psi psi}\\
&+\psi ^{(0)}\left(-\frac{d}{2}+s-1\right)-\left.\psi ^{(0)}\left(\frac{d}{2}\right)-\psi ^{(0)}\left(\frac{s}{2}\right)+\psi ^{(0)}(s-1)\right)
+{\cal O}\left(\frac{1}{N^2}\right)\,.\notag
\end{align}

We close this section by expanding the leading and next-to-leading
contributions to the $\langle \sigma\bar\psi\psi\rangle$ OPE
coefficients at $d=4-\epsilon_2$, $s=2-\epsilon_2/2+\epsilon_1$
to leading order in $\epsilon_{1,2}$,
\begin{equation}
\label{expanded result for C sigma psi psi}
\begin{aligned}
C_{\sigma\bar\psi\psi}^{(0)}|_{d=4} &= - \sqrt{\frac{2\epsilon_1}{N}}
+{\cal O}\left(\epsilon_{1,2}^\frac{3}{2},\,\frac{1}{N^\frac{3}{2}}\right)\,,\\
\delta C_{\sigma\bar\psi\psi} |_{d=4}&= -\frac{2}{N}+{\cal O}\left(\epsilon_{1,2},\,\frac{1}{N^2}\right)\,.
\end{aligned}
\end{equation}
These expressions will be compared with their
counterparts in the Gross-Neveu-Yukawa model in section~\ref{sec: LR GNY},
where we will perform perturbative $\epsilon$-expansion
near $d=4$, $s=2$, to the order ${\cal O}(\sqrt{\epsilon_{1,2}})$.

\subsection{Continuity of CFT data at $s=s_\star$}
\label{sec:continuity}

In this section we have derived various CFT data,
including scaling dimensions of the fields $\psi$, $\sigma$,
and the OPE coefficient $\langle\sigma\bar\psi\psi\rangle$
in the critical long-range Gross-Neveu model (\ref{HS form of LR GN})
at its interacting UV fixed point, working at the next-to-leading order in $1/N$ expansion. We are now going to check
continuity of this CFT data across the long-range--short-range
crossover point $s_\star$. This will serve as a non-trivial consistency
check for our results.

Recall that $s_\star$ is defined
as the threshold value of the long-range exponent parameter $s$,
above which the bi-local fermionic kinetic term (\ref{free generalized psi action}) is
irrelevant compared to the local (short-range) Dirac kinetic term. By calculating  scaling dimension of (\ref{free generalized psi action}) in the short-range model,
we obtain $s_\star = 2-2\gamma_{\hat\psi}$, where anomalous dimension
of fermionic field in the short-range critical Gross-Neveu model is
given by (\ref{gamma hat psi}). This immediately guarantees that the fermionic scaling dimension
is continuous,
\begin{equation}
[\psi]_{\textrm{UV}}|_{s=s_\star} = \frac{d-s_\star+1}{2}=\frac{d-1}{2}+\gamma_{\hat\psi}=[\hat \psi]_{\textrm{UV}}\,.
\end{equation}

Continuity of the full scaling dimension of the Hubbard-Stratonovich field
can be established by noticing that the difference between divergent graphs
contributing to the $\langle \sigma\sigma\rangle$ propagator in the long-range
model and the $\langle \hat\sigma\hat\sigma\rangle$ propagator in the short-range
model can be traced back to the vanishing anomalous dimension of the fermion in the former,
and therefore absence of the corresponding diagram, contributing $2\gamma_{\hat \psi}$.
This leads to the relation
\begin{equation}
\gamma_\sigma|_{s=2} =\gamma_{\hat\sigma}+ 2\gamma_{\hat \psi}\,,
\end{equation}
that can be verified explicitly using (\ref{gamma hat psi}), (\ref{gamma hat sigma}), (\ref{gamma sigma result}).
Here we have used $s_\star = 2 +{\cal O}(1/N)$, and expanded our expressions
to the next-to-leading order in $1/N$, that defined the regime of validity of our results.
We can then derive
\begin{equation}
[\sigma]_{\textrm{UV}}|_{s=s_\star} = s_\star -1+\gamma_\sigma|_{s=2} = 1 + (\gamma_\sigma|_{s=2}-2\gamma_{\hat \psi})
=[\hat\sigma]_{\textrm{UV}}\,.
\end{equation}

Finally, continuity of the $\langle\sigma\bar\psi\psi\rangle$ OPE coefficient
across the long-range--short-range crossover $s_\star$ can be established by
explicitly verifying validity of the following identities:\footnote{Analogous
consistency checks of continuity of CFT data across long-range--short-range
crossover in $O(N)$ vector models have recently been performed in \cite{Chai:2021arp}.}
\begin{equation}
\begin{aligned}
C_{\sigma\bar\psi\psi}^{(0)}|_{s=2} &= C_{\hat\sigma\bar{\hat\psi}\hat\psi}^{(0)} \,,\\
\left(C_{\sigma\bar\psi\psi}^{(0)}\,\delta C_{\sigma\bar\psi\psi}
-2\gamma_{\hat\psi}\frac{\partial C_{\sigma\bar\psi\psi}^{(0)}}{\partial s}\right)|_{s=2}
&=C_{\hat\sigma\bar{\hat\psi}\hat\psi}^{(0)}\,\delta C_{\hat\sigma\bar{\hat\psi}\hat\psi}\,.
\end{aligned}
\end{equation}

\section{Long-range Gross-Neveu-Yukawa model}
\label{sec: LR GNY}

In this section we will construct the long-range version of the
Gross-Neveu-Yukawa model \cite{Zinn-Justin:1991ksq}, describing
dynamics of the Dirac fermions $\theta^i$, $i=1,\dots,n$,
in the fundamental representation of the $U(n)$ symmetry group,
coupled to the pseudo-scalar $\phi$.
Our goal is to develop the model that flows to an IR-stable interacting fixed point
that is equivalent to the critical regime of the long-range Gross-Neveu model (\ref{HS form of LR GN}).

Establishing
 correspondence between the models begins
with the match of d.o.f.,
\begin{equation}
\psi^i\leftrightarrow\theta^i\,,\quad i = 1,\dots,n\,,\qquad\qquad \sigma\leftrightarrow\phi\,.
\end{equation}
The most non-trivial part of the proposed critical duality will manifest
itself in the equivalency of the CFT data of these two models. To this end, we start by noticing that
since the scaling dimension of the fermionic field $\psi^i$ in the long-range Gross-Neveu model
(\ref{HS form of LR GN}) is fixed exactly to (\ref{definition of Delta_psi}), we want to ensure
that the scaling dimension of the fermionic field $\theta^i$ in the long-range Gross-Neveu-Yukawa model
is similarly non-renormalized. This can be achieved by choosing
the bi-local fermionic kinetic term
\begin{equation}
\label{free generalized theta action}
S_{\textrm{LR\;Dirac}}[\theta] =C_f(s)\, \int d^dx\int d^dy\,\bar\theta^i(x)\gamma^\mu\theta^i(y)
\,\frac{(x-y)^\mu}{|x-y|^{d+s}}\,,
\end{equation}
for the fermion $\theta^i$.
The pre-factor $C_f(s)$ is given by (\ref{Cf(s)}),
ensuring that the momentum space propagator 
is canonically normalized. In position space this propagator is given by
\begin{equation}
\label{free theta prop}
\langle \theta(x)\bar\theta(0)\rangle = C_\theta\,\frac{\gamma^\mu\, x_\mu}{|x|^{2\Delta_\theta+1}}\,,
\end{equation}
where the scaling dimension is given by
\begin{equation}
\Delta_\theta = \frac{d-s+1}{2}\,,
\end{equation}
while the free propagator
amplitude is defined as
\begin{equation}
C_\theta = \frac{\Gamma\left(\frac{d-s}{2}+1\right)}{2^{s-1}\,\pi^\frac{d}{2}\,\Gamma\left(\frac{s}{2}\right)}\,.
\end{equation}

We then couple the model (\ref{free generalized theta action})
to the pseudo-scalar field $\phi$,
\begin{equation}
\label{LR GNY action}
S_{\textrm{LR\;GNY}}=S_{\textrm{LR\;Dirac}}[\theta] 
+\int d^dx\,\left(\frac{1}{2}\,(\partial\phi)^2 + g \,\phi\, \bar\theta\,\theta+\frac{h}{24}\,\phi^4\right)\,.
\end{equation}
Notice that analogously to the symmetry (\ref{Z2 sym of LR GN}) of the long-range Gross-Neveu model,
the action (\ref{LR GNY action}) enjoys the $\mathbb{Z}_2$ symmetry
\begin{equation}
\label{Z2 sym of LR GNY}
\begin{aligned}
&(x^1,\dots,x^{a-1},x^a,x^{a+1},\dots,x^d)\rightarrow (x^1,\dots,x^{a-1},-x^a,x^{a+1},\dots,x^d)\,,\\
&\phi\rightarrow-\phi\,,\quad\theta\rightarrow \gamma^a\theta\,,\quad
\bar\theta\rightarrow-\bar\theta\gamma^a\,,
\end{aligned}
\end{equation}
defined for any given $a=1,\dots,d$. Additionally, since we can redefine $\phi\rightarrow-\phi$, this implies that every non-trivial
fixed point value $g_\star$ is paired up with a physically equivalent fixed point $-g_\star$.

In the free regime, the propagator of the scalar field $\phi$ is given by
\begin{equation}
\label{free <phi phi>}
\langle \phi(x)\phi(0)\rangle = \frac{C_\phi}{|x|^{2\Delta_\phi}}\,,
\end{equation}
where the scaling dimension is given by
\begin{equation}
\Delta_\phi = \frac{d}{2} - 1\,,
\end{equation}
while the free propagator amplitude is
\begin{equation}
C_\phi = \frac{\Gamma\left(\frac{d}{2}-1\right)}{4\pi^\frac{d}{2}}\,.
\end{equation}
Near the free regime, $g=0$, $h=0$, the coupling constants
therefore have dimensions
\begin{equation}
[g]_{\textrm{free}} = s-\frac{d}{2}\,,\qquad
[h]_{\textrm{free}} = 4-d\,.
\end{equation}
Choosing
\begin{equation}
\label{d and s in terms of epsilon12}
d = 4-\epsilon_2\,,\qquad s = \frac{d}{2} + \epsilon_1 = 2-\frac{\epsilon_2}{2}+\epsilon_1\,,
\end{equation}
we can make the couplings $g$, $h$ slightly relevant near the UV.
We then have
\begin{equation}
\label{Delta theta and phi in terms of epsilon12}
\Delta_\theta = \frac{3}{2} - \frac{\epsilon_1}{2} - \frac{\epsilon_2}{4}\,,\qquad
\Delta_\phi = 1 - \frac{\epsilon_2}{2}\,.
\end{equation}
Let us now redefine the coupling constants to be dimensionless, 
\begin{equation}
\label{LR GNY action redef gh}
S_{\textrm{LR\;GNY}}=S_{\textrm{LR\;Dirac}}[\theta] 
+\int d^dx\,\left(\frac{1}{2}\,(\partial\phi)^2 + g \,\mu^{\epsilon_1}\,\phi\, \bar\theta\,\theta+\frac{h\,\mu^{\epsilon_2}}
{24}\,\phi^4\right)\,,
\end{equation}
where $\mu$ is an arbitrary RG scale.
The relevant interaction in the UV activates
an RG flow, that brings the theory to an interacting Wilson-Fisher fixed point $(g_\star, h_\star)$
in the IR, that can be found perturbatively using $\epsilon$-expansion.
Our goal in this section is to find this fixed point and study the corresponding critical theory.

While we will carry out our calculation at the next-to-leading order in $\epsilon$-expansion,
our results in this section are going to be exact in $1/N$. However, since we are largely
motivated by the desire to compare the corresponding critical theory at the fixed
point with the critical long-range Gross-Neveu model, that we discussed in section~\ref{sec:LR GN},
we will subsequently perform algebraic expansion of our results in $1/N$. Specifically,
the fixed-point values of the dimensionless coupling constants can be represented as
\begin{equation}
\label{1/N expansion of g and h in GNY}
\begin{aligned}
g_\star &= g_0\,\left(1+\frac{g_1}{N} + {\cal O}\left(\frac{1}{N^2}\right)\right)\,,\\
h_\star &= h_0\,\left(1+\frac{h_1}{N} + {\cal O}\left(\frac{1}{N^2}\right)\right)\,.
\end{aligned}
\end{equation}
From the action (\ref{LR GNY action redef gh}) it is clear that $\theta ^ i\sim{\cal O}(1/N^0)$,
$\phi\sim{\cal O}(\sqrt{N})$, and therefore
\begin{equation}
g_0\sim {\cal O}\left(\frac{1}{\sqrt{N}}\right)\,,\qquad
h_0\sim {\cal O}\left(\frac{1}{N}\right)\,.
\end{equation}
Using the Callan-Symanzik equation
\begin{equation}
\left(\mu\frac{\partial}{\partial\mu} + \beta_g\frac{\partial}{\partial g} + \gamma_\phi\right) \langle
 \phi \bar\theta\theta\rangle = 0\,,
\end{equation}
where we took into account $\gamma_\theta = 0$, and expanded to the leading
order in $1/N$, we then obtain
\begin{equation}
\gamma_\phi = \epsilon_1 + {\cal O}\left(\frac{1}{N}\right)\,,
\end{equation}
and as a result using (\ref{d and s in terms of epsilon12}),
(\ref{Delta theta and phi in terms of epsilon12}) we obtain $[\phi]_{\textrm{IR}} = s-1+ {\cal O}\left(\frac{1}{N}\right)$.
While below in this section we will calculate $\gamma_\phi$ at leading order in $\epsilon$-expansion
but to all orders in $1/N$, we can already notice that such a large-$N$ behavior of $[\phi]_{\textrm{IR}}$
matches the scaling dimension (\ref{[sigma] UV}) of the Hubbard-Stratonovich field
of the critical long-range Gross-Neveu model. Moreover, it indicates that the quartic
and the local kinetic term for the field $\phi$ in the action (\ref{LR GNY action redef gh}),
that naively distinguish it from the long-range Gross-Neveu model (\ref{HS form of LR GN}) in the UV,
are in fact irrelevant in the IR fixed point of the model (\ref{LR GNY action redef gh}).

We will carry out perturbative $\epsilon$-expansion in position space.
The Feynman rules that we are going to be using  in this section for 
the free fermionic and scalar propagator, and the cubic interaction vertex,
are given by
\begin{center}
  \begin{picture}(452,62) (9,50)
    \SetWidth{1.0}
    \SetColor{Black}
    \Vertex(30,104){2}
    \Line[arrow,arrowpos=0.5,arrowlength=5,arrowwidth=2,arrowinset=0.2](30,104)(132,104)
    \Vertex(132,104){2}
    \Text(145,104)[lb]{\scalebox{0.8}{$x$}}
    \Text(172,94)[lb]{\scalebox{1}{$=C_\theta\,\frac{x^\mu\gamma_\mu}{|x|^{2\Delta_\theta+1}}$}}
    \Text(15,104)[lb]{\scalebox{0.8}{$0$}}
    \Vertex(30,64){2}
    \Line[](30,64)(132,64)
    \Vertex(132,64){2}
    \Text(145,64)[lb]{\scalebox{0.8}{$x$}}
    \Text(172,54)[lb]{\scalebox{1}{$=C_\phi\,\frac{1}{|x|^{2\Delta_\phi}}$}}
    \Text(15,64)[lb]{\scalebox{0.8}{$0$}}
    \Line[arrow,arrowpos=0.5,arrowlength=5,arrowwidth=2,arrowinset=0.2](348,84)(366,100)
    \Line[arrow,arrowpos=0.5,arrowlength=5,arrowwidth=2,arrowinset=0.2](366,66)(348,84)
    \Line[](348,84)(325,84)
    \Vertex(348,84){4}
    \Text(400,77)[lb]{\scalebox{1}{$=-g$}}
  \end{picture}
\end{center}
For fermionic and scalar lines with explicitly specified exponents we will
define the corresponding amplitudes to be equal to unity.
Together with the quartic interaction vertex, the corresponding
Feynman rules are given by
\begin{center}
  \begin{picture}(452,62) (9,50)
    \SetWidth{1.0}
    \SetColor{Black}
    \Vertex(30,104){2}
    \Line[arrow,arrowpos=0.5,arrowlength=5,arrowwidth=2,arrowinset=0.2](30,104)(132,104)
    \Vertex(132,104){2}
    \Text(145,104)[lb]{\scalebox{0.8}{$x$}}
    \Text(172,94)[lb]{\scalebox{1}{$=\frac{x^\mu\gamma_\mu}{|x|^{2\Delta+1}}$}}
    \Text(15,104)[lb]{\scalebox{0.8}{$0$}}
    \Text(75,114)[lb]{\scalebox{0.8}{$2\Delta$}}
    \Vertex(30,64){2}
    \Line[](30,64)(132,64)
    \Vertex(132,64){2}
    \Text(145,64)[lb]{\scalebox{0.8}{$x$}}
    \Text(172,54)[lb]{\scalebox{1}{$=\frac{1}{|x|^{2\Delta}}$}}
    \Text(15,64)[lb]{\scalebox{0.8}{$0$}}
    \Text(75,74)[lb]{\scalebox{0.8}{$2\Delta$}}
    \Line[](332,68)(366,100)
    \Line[](366,66)(332,100)
    \Vertex(348,84){4}
    \Text(400,77)[lb]{\scalebox{1}{$=-h$}}
  \end{picture}
\end{center}

\subsection{$\langle\phi\phi\rangle$}
\label{sec: phi phi}

At the leading order in $\epsilon$-expansion, the $\langle\phi\phi\rangle$
propagator is given by the free field expression (\ref{free <phi phi>}).
One-loop correction to this expression is determined by the following diagram:\footnote{
We argue that the two-loop ${\cal O}(h^2)$ diagram is of a sub-leading order ${\cal O}(\epsilon_1^2)$.
This is analogous to the short-range Gross-Neveu-Yukawa model, where at the
Wilson-Fisher fixed point in $d=4-\epsilon$ dimensions
the leading order couplings behave as $g={\cal O}(\sqrt{\epsilon})$, 
$h={\cal O}(\epsilon)$.}
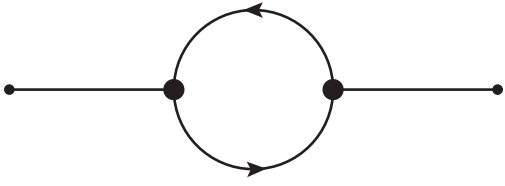
\begin{center}
  \begin{picture}(296,70) (-20,10)
    \SetWidth{1.0}
    \SetColor{Black}
    \Arc[arrow,arrowpos=0.5,arrowlength=5,arrowwidth=2,arrowinset=0.2](130,44)(30,180,364)
    \Arc[arrow,arrowpos=0.5,arrowlength=5,arrowwidth=2,arrowinset=0.2](130,44)(30,0,180)
    \Line[](38,44)(101,44)
    \Line[](160,44)(224,44)
    \Vertex(100,44){4.001}
    \Vertex(160,44){4}
    \Vertex(38,44){2}
    \Vertex(222,44){2.001}
  \end{picture}
\end{center}
In this diagram, the Feynman rule for the fermionic loop contributes
the factor of $-1$. Combining two fermionic propagators forming
a loop renders another factor of $-1$, as we reviewed in Appendix~\ref{app_a}.
Using the propagator merging relations, and combining the resulting
expression with the leading order propagator (\ref{free <phi phi>}), we obtain
\begin{equation}
\begin{aligned}
\langle\phi(x)\phi(0)\rangle &= \frac{C_\phi}{|x|^{2-\epsilon_2}}\,
\left(1+g_\star^2NC_\theta^2C_\phi U\left(1-\frac{\epsilon_2}{2},
3-\epsilon_1-\frac{\epsilon_2}{2},\epsilon_1\right)\right.\\
&\times\left.U\left(1-\frac{\epsilon_2}{2},
2-\epsilon_1-\frac{\epsilon_2}{2},1+\epsilon_1\right)(\mu |x|)^{2\epsilon_1}\right)\\
&\rightarrow \frac{C_\phi}{|x|^{2-\epsilon_2}}\,\left(1-\frac{N\,g_\star^2}{16\pi^2}\,\log(\mu |x|)\right)\,,
\end{aligned}
\end{equation}
where in the last line we performed expansion in $\epsilon_{1}$,  collected leading order
terms in front of the logarithm, and discarded the $1/\epsilon_1$ term.\footnote{These pure divergencies
can be absorbed into the wave-function 
\begin{equation}
\phi\rightarrow\sqrt{1+\frac{\gamma_\phi}{\epsilon_1}}\,\phi\,.
\end{equation}
We keep 
renormalizations implicit here and everywhere else in this section.}
Consequently,  anomalous dimension of the field $\phi$ at the IR f.p. is given by
\begin{equation}
\label{gamma phi general}
\gamma_\phi = \frac{N\,g_\star^2}{32\pi^2} + {\cal O}(\epsilon_1^2)\,.
\end{equation}

\subsection{$\langle \phi\bar\theta\theta\rangle$}
\label{sec: phi theta theta}

We now proceed to calculation of the three-point function $\langle \phi\bar\theta\theta\rangle$
up to the one-loop order.
At the leading order it is given by the diagram
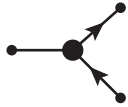
\begin{center}
  \begin{picture}(652,22) (140,67)
    \SetWidth{1.0}
    \SetColor{Black}
    \Vertex(366,100){2}
    \Vertex(366,66){2}
    \Vertex(325,84){2}
    \Line[arrow,arrowpos=0.5,arrowlength=5,arrowwidth=2,arrowinset=0.2](348,84)(366,100)
    \Line[arrow,arrowpos=0.5,arrowlength=5,arrowwidth=2,arrowinset=0.2](366,66)(348,84)
    \Line[](348,84)(325,84)
    \Vertex(348,84){4}
  \end{picture}
\end{center}
The corresponding contribution to the three-point function reads\footnote{At the leading order
in $\epsilon$-expansion one drops ${\cal O}(\epsilon)$ contributions to the exponents of $|x_{ij}|$.}
\begin{equation}
\label{general result for phi theta theta}
\langle \phi (x_1)\bar\theta(x_2)\theta (x_3)\rangle\Bigg|_{\textrm{normalized}}
=\frac{C_{\phi\bar\theta\theta}\,x_{12}^\mu\gamma_\mu\,x_{31}^\nu\gamma_\mu}
{(|x_{12}||x_{13}||x_{23}|)^2}\,,
\end{equation}
where
\begin{equation}
\label{leading phi theta theta amplitude}
C_{\phi\bar\theta\theta} = - g_\star\, C_\phi^\frac{1}{2} \,C_\theta\,\pi^\frac{d}{2}\,A(1)\,
V\left(\frac{d-1}{2},\frac{d-1}{2}\right)\Bigg|_{d=4,s=2}+{\cal O}(g_\star^3)\,.
\end{equation}
The motivation behind such a choice
is that it provides us with universal conventions allowing us to compare
the corresponding OPE coefficients in different models.

At the next-to-leading order in $\epsilon$-expansion we have a contribution from 
the following vertex correction one-loop diagram:
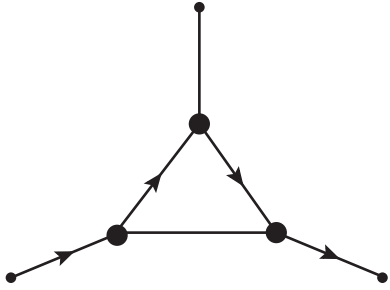
\begin{center}
  \begin{picture}(163,114) (0,-10)
    \SetWidth{1.0}
    \SetColor{Black}
    \Line[](81,95)(81,51)
    \Line[](50,10)(110,10)
    \Line[arrow,arrowpos=0.5,arrowlength=5,arrowwidth=2,arrowinset=0.2](10,-7)(51,10)
    \Line[arrow,arrowpos=0.5,arrowlength=5,arrowwidth=2,arrowinset=0.2](110,10)(151,-7)
    \Line[arrow,arrowpos=0.5,arrowlength=5,arrowwidth=2,arrowinset=0.2](48,8)(81,51)
    \Line[arrow,arrowpos=0.5,arrowlength=5,arrowwidth=2,arrowinset=0.2](81,51)(110,10)
    \Vertex(50,9){4}
    \Vertex(81,51){4}
    \Vertex(110,10){4}
    \Vertex(10,-7){2}
    \Vertex(150,-7){2}
    \Vertex(81,95){2}
  \end{picture}
\end{center}
The corresponding term in the three-point function is given by
\begin{equation}
\langle \phi (x_1)\bar\theta(x_2)\theta (x_3)\rangle\Bigg|_{\textrm{normalized}}
=(-g_\star)^3\,C_\phi^\frac{3}{2}\,C_\theta^3\,V(\mu_0)\,,
\end{equation}
where $V(\mu_0)$ is determined by the regularized diagram 
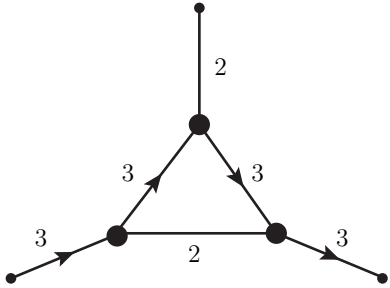
\begin{center}
  \begin{picture}(163,114) (0,-10)
    \SetWidth{1.0}
    \SetColor{Black}
    \Line[](81,95)(81,51)
    \Line[](50,10)(110,10)
    \Line[arrow,arrowpos=0.5,arrowlength=5,arrowwidth=2,arrowinset=0.2](10,-7)(51,10)
    \Line[arrow,arrowpos=0.5,arrowlength=5,arrowwidth=2,arrowinset=0.2](110,10)(151,-7)
    \Line[arrow,arrowpos=0.5,arrowlength=5,arrowwidth=2,arrowinset=0.2](48,8)(81,51)
    \Line[arrow,arrowpos=0.5,arrowlength=5,arrowwidth=2,arrowinset=0.2](81,51)(110,10)
    \Vertex(50,9){4}
    \Vertex(81,51){4}
    \Vertex(110,10){4}
    \Vertex(10,-7){2}
    \Vertex(150,-7){2}
    \Vertex(81,95){2}
    \Text(87,70)[lb]{\scalebox{0.8}{$2$}}
    \Text(52,30)[lb]{\scalebox{0.8}{$3$}}
    \Text(101,30)[lb]{\scalebox{0.8}{$3$}}
    \Text(77,-1)[lb]{\scalebox{0.8}{$2$}}
    \Text(19,5)[lb]{\scalebox{0.8}{$3$}}
    \Text(133,5)[lb]{\scalebox{0.8}{$3$}}
  \end{picture}
\end{center}
This diagram is  logarithmically divergent. It can be regularized
by imposing the UV cut-off $|x|\geq 1/\mu_0$, and subtracting the divergent term $\log(\mu_0)$
with the vertex counterterm,  at the cost of introducing an RG scale $\mu$:
\begin{equation}
\label{regularized log integral}
\int _{1/\mu_0}\frac{d|x|}{|x|}\Bigg|_{\textrm{regularized}} = \log(\mu)\,.
\end{equation}
The integrals over the other two vertices are finite, and can be taken 
using the star-triangle relation, reviewed in Appendix~\ref{app_a}. We then arrive at
\begin{equation}
V(\mu) = - 2\pi^4 \, A(1)\,V\left(\frac{3}{2},\frac{3}{2}\right)\, U\left(1,\frac{3}{2},\frac{3}{2}\right)\Bigg|_{d=4,s=2}
\,\frac{x_{12}^\mu\gamma_\mu\,x_{31}^\nu\gamma_\mu}
{(|x_{12}||x_{13}||x_{23}|)^2}\,
\log(\mu)\,.
\end{equation}

Finally, dressing the $\phi$ leg of the tree-level $\langle \phi\bar\theta\theta\rangle$ diagram, we obtain 
\begin{equation}
\langle \phi (x_1)\bar\theta(x_2)\theta (x_3)\rangle\Bigg|_{\textrm{normalized}}
=-2\gamma_\phi\,\log(\mu)\,\frac{C_{\phi\bar\theta\theta}\,x_{12}^\mu\gamma_\mu\,x_{31}^\nu\gamma_\mu}
{(|x_{12}||x_{13}||x_{23}|)^2}\,,
\end{equation}

Using the Callan-Symanzik equation
\begin{equation}
\left(\mu\frac{\partial}{\partial\mu} + \beta_g\frac{\partial}{\partial g} + \gamma_\phi\right) \langle
 \phi (x_1)\bar\theta(x_2)\theta (x_3)\rangle = 0
\end{equation}
we then obtain
\begin{equation}
\beta_g = -\epsilon_1\,g +\frac{N+4}{32\pi^2}\,g^3\,.
\end{equation}
A non-trivial IR fixed point is then achieved at\footnote{Recall that $-g_\star$ is a physically equivalent fixed point.}
\begin{equation}
\label{g star}
g_\star = 4\pi\,\sqrt{\frac{2\epsilon_1}{N+4}}\,.
\end{equation}
Expanding (\ref{g star}) in $1/N$ and plugging it into
 (\ref{leading phi theta theta amplitude}), we then obtain the normalized three-point
function amplitude 
\begin{equation}
\label{C phi theta theta 1/N}
C_{\phi\bar\theta\theta} = -\sqrt{\frac{2\epsilon_1}{N}}\,
\left(1-\frac{2}{N}+{\cal O}\left(\frac{1}{N^2}\right)\right) + {\cal O}(\epsilon_1^{3/2})\,.
\end{equation}

On the other hand, substituting (\ref{g star})
into (\ref{gamma phi general}) and expanding in $1/N$, we obtain
\begin{equation}
\label{[phi] 1/N}
[\phi] _{\textrm{IR}} = \Delta_\phi + \gamma_\phi = 1-\frac{\epsilon_2}{2} + \epsilon_1 - \frac{4\epsilon_1}{N}
+{\cal O}\left(\epsilon_1^2,\frac{1}{N^2}\right)
=s-1 - \frac{4\epsilon_1}{N} +{\cal O}\left(\epsilon_1^2,\frac{1}{N^2}\right)\,,
\end{equation}
where we used (\ref{d and s in terms of epsilon12}), (\ref{Delta theta and phi in terms of epsilon12}).

Notice that while our calculation is perturbative in $\epsilon_{1,2}$, it is exact in $1/N$.
However, performing algebraic $1/N$ expansion allows us to compare the resulting
expressions (\ref{C phi theta theta 1/N}), (\ref{[phi] 1/N})
at the next-to-leading order in $1/N$ with their counterparts (\ref{expanded result for C sigma psi psi}),
(\ref{gamma sigma expanded result})
in the large-$N$ long-range Gross-Neveu model, revealing an exact agreement.

These non-trivial matches of CFT data at the next-to-leading order in $1/N$
expansion provide a strong supportive evidence in favor of the statement
that the UV fixed point of the long-range Gross-Neveu model,
and the IR fixed point of the long-range Gross-Neveu-Yukawa model
are in fact described by the same CFT. While our calculations
have been performed in $\epsilon$-expansion near $d=4$, $s=2$ for the Gross-Neveu-Yukawa model,
and in $1/N$ expansion for the Gross-Neveu model, we suggest that this critical universality
in fact holds true for the entire domain (\ref{bounds on s}), (\ref{bounds on d}).

\subsection{$\langle\phi\phi\phi\phi\rangle$}
\label{sec: phi phi phi phi}

In this section we are going to use the four-point function 
$\langle\phi\phi\phi\phi\rangle$ at the next-to-leading order in $\epsilon$-expansion to calculate
the fixed-point value of the coupling $h_\star$.
The leading order contribution to the normalized four-point function is given by the tree-level diagram
\begin{center}
  \begin{picture}(452,62) (200,50)
    \SetWidth{1.0}
    \SetColor{Black}
    \Text(200,68)[lb]{\scalebox{1}{$\langle\phi\phi\phi\phi\rangle\Bigg|_{\textrm{normalized}}\supset$}}
    \Text(400,78)[lb]{\scalebox{1}{$=-h\,C_\phi^2\;\;\times$}}
    \Text(550,78)[lb]{\scalebox{1}{$\equiv-h\,C_\phi^2\,{\cal V}$}}
    \Vertex(332,68){2}
    \Vertex(366,100){2}
    \Vertex(366,66){2}
    \Vertex(332,100){2}
    \Line[](332,68)(366,100)
    \Line[](366,66)(332,100)
    \Vertex(348,84){4}
    \Vertex(482,68){2}
    \Vertex(516,100){2}
    \Vertex(516,66){2}
    \Vertex(482,100){2}
    \Line[](482,68)(516,100)
    \Line[](516,66)(482,100)
    \Vertex(498,84){4}
    \Text(503,94)[lb]{\scalebox{0.8}{$2$}}
    \Text(486,85)[lb]{\scalebox{0.8}{$2$}}
    \Text(510,77)[lb]{\scalebox{0.8}{$2$}}
    \Text(494,70)[lb]{\scalebox{0.8}{$2$}}
  \end{picture}
\end{center}
While the four-line vertex on the r.h.s. of this diagrammatic
equation is conformal, we do not need to take this integral
for our purposes. Instead, we denote it as ${\cal V}$
and keep it implicit. 

At the next-to-leading order we have contributions due
to the dressed $\phi$ legs of the tree-level diagram, as 
well as two vertex corrections. The former is given by
\begin{equation}
\langle\phi\phi\phi\phi\rangle\Bigg|_{\textrm{normalized}}\supset
4\times (-2\gamma_\phi)\,(-h)\, C_\phi^2\,  {\cal V}\,,
\end{equation}
One of the vertex corrections is determined by the diagram
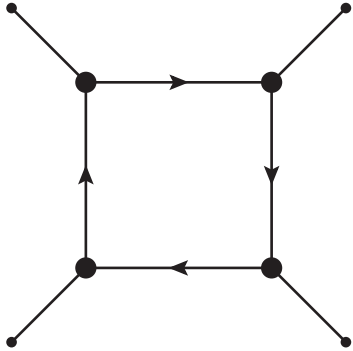
\begin{center}
  \begin{picture}(142,142) (20,-13)
    \SetWidth{1.0}
    \SetColor{Black}
    \Line[arrow,arrowpos=0.5,arrowlength=5,arrowwidth=2,arrowinset=0.2](56,93)(126,93)
    \Line[arrow,arrowpos=0.5,arrowlength=5,arrowwidth=2,arrowinset=0.2](126,93)(126,23)
    \Line[arrow,arrowpos=0.5,arrowlength=5,arrowwidth=2,arrowinset=0.2](126,23)(56,23)
    \Line[arrow,arrowpos=0.5,arrowlength=5,arrowwidth=2,arrowinset=0.2](56,23)(56,93)
    \Line[](126,93)(154,121)
    \Line[](126,23)(154,-5)
    \Line[](56,23)(28,-5)
    \Line[](56,93)(28,121)
    \Vertex(28,121){2}
    \Vertex(154,121){2}
    \Vertex(154,-5){2}
    \Vertex(28,-5){2}
    \Vertex(56,23){4}
    \Vertex(126,23){4}
    \Vertex(126,93){4}
    \Vertex(56,93){4}
  \end{picture}
\end{center}
Two of its vertices can be calculated using the uniqueness relation reviewed in Appendix~\ref{app_a}.
The third integral is logarithmically divergent, and can be regularized using the UV cut-off,
with the divergence being subsequently subtracted using vertex counter-term, resulting in (\ref{regularized log integral}). 
This way the diagram is reduced to ${\cal V}$, that implicitly contains the fourth integral.
Assembling everything together we obtain (here 6 is the symmetry factor and overall minus sign
is due to the fermionic loop)
\begin{equation}
\langle\phi\phi\phi\phi\rangle\Bigg|_{\textrm{normalized}}\supset  -6Nh^4C_\phi^2C_\theta^4\,
\left(\pi^\frac{d}{2}A(1)V\left(\frac{3}{2},\frac{3}{2}\right)\right)^2\Bigg|_{d=4,s=2}\,2\pi^2\,\log(\mu)\,{\cal V}\,.
\end{equation}
The other vertex correction diagram contributing to $\langle\phi\phi\phi\phi\rangle$ 
at the next-to-leading order in $\epsilon$-expansion is given by
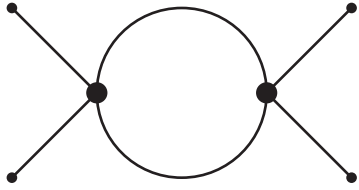
\begin{center}
  \begin{picture}(146,82) (15,-15)
    \SetWidth{1.0}
    \SetColor{Black}
    \Line[](24,58)(56,26)
    \Line[](56,26)(24,-6)
    \Arc[clock](88,26)(32,-180,-360)
    \Arc[](88,26)(32,-180,0)
    \Line[](120,26)(152,58)
    \Line[](120,26)(152,-6)
    \Vertex(24,58){2}
    \Vertex(24,-6){2}
    \Vertex(56,26){4}
    \Vertex(120,26){4}
    \Vertex(152,58){2}
    \Vertex(152,-6){2}
  \end{picture}
\end{center}
The corresponding contribution has the form  (here $3/2$ is the symmetry factor)
\begin{equation}
\langle\phi\phi\phi\phi\rangle\Bigg|_{\textrm{normalized}}\supset  \frac{3}{2}\,g^2\,C_\phi^4\,2\pi^2\,\log(\mu)\,{\cal V}\,,
\end{equation}
where we regularized one of the integrals.
The common factor ${\cal V}$ cancels out from the Callan-Symanzik equation,
\begin{equation}
\left( \mu\frac{\partial}{\partial\mu} + \beta_h\frac{\partial}{\partial h}
+4\gamma_\phi \right)\langle\phi\phi\phi\phi\rangle = 0\,,
\end{equation}
resulting in 
\begin{equation}
\beta_h = -\epsilon_2 h+\frac{2 N g^2\,( h - 6 g^2)  +3 h^2}{16 \pi ^2}\,.
\end{equation}
The corresponding IR stable fixed point is given by
\begin{equation}
h_\star = \frac{8}{3} \pi ^2 \left(\sqrt{\frac{16 N (N+36)
\epsilon_1^2}{(N+4)^2}-\frac{8 N \epsilon_1 \epsilon_2}{N+4}+\epsilon_2^2}
-\frac{4 N \epsilon_1}{N+4}+\epsilon_2\right)\,.
\end{equation}

\section{Discussion}
\label{sec:discussion}

In this paper we have constructed new long-range Gross-Neveu and Gross-Neveu-Yukawa
fermionic models. For the choice
of parameters $d/2<s<\min(d/2+1,s_\star)$, $1<d<4$,
we demonstrated that these
models flow to critical regimes (in the UV and the IR respectively), and matched CFT data of these models within
the overlapping regimes of validity. Our paper therefore furnishes a non-trivial
generalization of the well-known critical universality between the short-range
Gross-Neveu and Gross-Neveu-Yukawa models. Moreover, the long-range critical
universality established in this paper complements the recently
observed critical universality between the UV fixed point of the long-range $O(N)$
vector model, and the IR fixed point of the long-range $O(N)$ scalar multiplet coupled to
a fundamental scalar singlet with cubic self-interaction.

Conformal structure of the three-point function $\langle \phi\bar\theta\theta\rangle$
in the long-range Gross-Neveu-Yukawa model indicates that its fixed
point enjoys the full conformal symmetry, at least at the order ${\cal O}(\sqrt{\epsilon_{1,2}})$
at which we have performed explicit calculations.
Additionally,  one can check the two-point functions
that must vanish if the full conformal symmetry is valid. For instance, the two-point function
$\langle \phi(x_1)\bar\psi\psi(x_2)\rangle$
vanishes up to a contact term, and the correlator $\langle \phi ^2(x_1)\phi^4(x_2)\rangle$
is  zero at the considered order in $\epsilon$-expansion.
Due to the critical duality argument,
provided in this paper, this implies that the critical long-range Gross-Neveu model
is also in fact described by a CFT. It would be interesting to see how conformal symmetry at the
long-range fixed point of the considered fermionic models can be established 
analogously to the arguments of \cite{Paulos:2015jfa,Behan:2017dwr,Behan:2017emf,Behan:2018hfx,Giuliani:2020aot}.

As we reviewed in section~\ref{sec:intro}, the long-range critical vector model can be
found at the end of an RG flow, triggered by coupling the critical short-range
vector model to a generalized free field of dimension $(d+s)/2$ \cite{Behan:2017dwr,Behan:2017emf}.
We suggest that similarly one can couple the short-range Gross-Neveu model
to a long-range fermion $\chi$ of dimension $\Delta_{\chi} = (d+s-1)/2$,
\begin{equation}
\hat S = S_{\textrm{LR Dirac}}[\chi]|_{2-s} + \lambda\,\int d^dx\,
(\hat{\bar\psi}\chi + \bar\chi\hat\psi)\,.
\end{equation} 
Near $s=s_\star$ we therefore obtain $[\lambda] = s_\star - s$, where 
$s_\star = 2-2\gamma_{\hat\psi}$ defines the long-range--short-range
crossover threshold. For small $\delta = s_\star - s$ one can therefore study
the model perturbatively in $\lambda$.

Long-range vector models have recently been used
in systems exhibiting persistent symmetry breaking \cite{Chai:2021djc}.\footnote{See
also \cite{Chai:2020zgq,Chai:2020hnu} for recent work on persistent symmetry breaking
in vector models.} While these systems have so far been considered to be exclusively bosonic,
it would be interesting to see what role the long-range
fermions play in models manifesting persistent symmetry breaking.

\section*{Acknowledgements} \noindent  
The work of N.C. and R.S. is partially supported by the Binational Science Foundation (grant No. 2016186), the Israeli Science Foundation Center of Excellence (grant No. 2289/18), and by the Quantum Universe I-CORE program of the Israel Planning and Budgeting Committee (grant No. 1937/12). The work of N.C. is partially supported by Yuri Milner scholarship.
The work of S.C. is supported by the Infosys Endowment for the study of the Quantum Structure of Spacetime. 
R.S. would like to thank CPHT at Ecole Polytechnique, France for hospitality during the course of this work.
The work of M.G. is supported by DOE grant DE-SC0011842.

\appendix

\section{Some useful identities}
\label{app_a}

In this appendix we collect some known expressions and identities, that are useful
to carry out perturbation theory calculations in position space \cite{DEramo:1971hnd,Symanzik:1972wj,Gracey:1990wi}
(see also \cite{Preti:2018vog} for a recent review).

Loop diagram in position space are simply additive:
\begin{center}
  \begin{picture}(257,50) (0,0)
    \SetWidth{1.0}
    \SetColor{Black}
    \Arc[clock](81,-39)(77.006,127.614,52.38600001)
    \Arc[](80,86)(80,-126.87,-53.13)
    \Line[](160,22)(256,22)
    \Vertex(160,22){2.005}
    \Vertex(256,22){2.005}
    \Vertex(33,22){2.002}
    \Vertex(129,22){2.003}
    \Text(142,20)[lb]{$=$}
    \Text(77,-5)[lb]{\scalebox{0.8}{$2b$}}
    \Text(77,43)[lb]{\scalebox{0.8}{$2a$}}
    \Text(190,27)[lb]{\scalebox{0.8}{$2(a+b)$}}
  \end{picture}
\end{center}
When fermions are involved, we obtain
\begin{center}
\scalebox{0.9}{
  \begin{picture}(197,71) (39,-5)
    \SetWidth{1.0}
    \SetColor{Black}
    \Vertex(40,31){2.001}
    \Arc[arrow,arrowpos=0.5,arrowlength=5,arrowwidth=2,arrowinset=0.2,clock](82.5,-4)(55.057,141,39)
    \Vertex(125,31){2}
    \Arc[clock](82.5,65)(55.057,-39,-141)
    \Text(135,28)[lb]{$=$}
    \Vertex(155,31){2}
    \Line[arrow,arrowpos=0.5,arrowlength=5,arrowwidth=2,arrowinset=0.2](155,31)(215,31)
    \Vertex(215,31){2.001}
    \Text(165,37)[lb]{\scalebox{0.8}{$2\Delta_1+2\Delta_2$}}
    \Text(80,57)[lb]{\scalebox{0.8}{$2\Delta_1$}}
    \Text(80,-3)[lb]{\scalebox{0.8}{$2\Delta_2$}}
  \end{picture}
  }
\end{center}
\begin{center}
\scalebox{0.9}{
  \begin{picture}(197,71) (39,-5)
    \SetWidth{1.0}
    \SetColor{Black}
    \Vertex(40,31){2.001}
    \Arc[arrow,arrowpos=0.5,arrowlength=5,arrowwidth=2,arrowinset=0.2,clock](82.5,-4)(55.057,141,39)
    \Vertex(125,31){2}
    \Arc[arrow,arrowpos=0.5,arrowlength=5,arrowwidth=2,arrowinset=0.2,clock](82.5,65)(55.057,-39,-141)
    \Text(135,28)[lb]{$=$}
    \Vertex(155,31){2}
    \Line[](155,31)(215,31)
    \Vertex(215,31){2.001}
    \Text(165,37)[lb]{\scalebox{0.8}{$2\Delta_1+2\Delta_2$}}
    \Text(80,57)[lb]{\scalebox{0.8}{$2\Delta_1$}}
    \Text(80,-3)[lb]{\scalebox{0.8}{$2\Delta_2$}}
    \Text(225,27)[lb]{$\times\;\; (-\mathbb{I})$}
  \end{picture}
  }
\end{center}

The propagator merging relation identity is given by
\begin{equation}
\label{prop merging}
\int d^d x_2\, \frac{1}{|x_2|^{2a}|x_1-x_2|^{2b}}
=U(a,b,d-a-b)\,\,\frac{1}{|x_{1}|^{2a+2b-d}}\,,
\end{equation}
where we defined
\begin{align}
\label{U def}
U(a,b,c) \,=\, \pi^\frac{d}{2} A(a)A(b)A(c)\,.
\end{align}
Here we have introduced
\begin{equation}
A(x) \,=\, \frac{\Gamma\left(\frac{d}{2}-x\right)}{\Gamma(x)}\,.
\end{equation}
This relation can also be represented diagrammatically as
\begin{center}
  \begin{picture}(98,10) (130,-60)
    \SetWidth{1.0}
    \SetColor{Black}
    \Line[](30,-58)(90,-58)
    \Line[](90,-58)(150,-58)
    \Line[](180,-58)(240,-58)
    \Vertex(30,-58){2.0001}
    \Vertex(90,-58){4.0001}
    \Vertex(150,-58){2.000001}
    \Vertex(180,-58){2.0001}
    \Vertex(240,-58){2.000001}
    \Text(55,-53)[lb]{\scalebox{0.801}{$2a$}}
    \Text(115,-53)[lb]{\scalebox{0.801}{$2b$}}
    \Text(163,-61)[lb]{$=$}
    \Text(180,-53)[lb]{\scalebox{0.801}{$2(a+b)-d$}}
    \Text(250,-63)[lb]{$\times U(a,b,d-a-b)$}
  \end{picture}
\end{center}

Propagator merging relations with fermions are given by
\begin{center}
\scalebox{0.9}{
  \begin{picture}(271,37) (49,-10)
    \SetWidth{1.0}
    \SetColor{Black}
    \Vertex(-20,7){2.001}
    \Line[arrow,arrowpos=0.5,arrowlength=5,arrowwidth=2,arrowinset=0.2](-20,7)(35,7)
    \Line[](35,7)(85,7)
    \Vertex(85,7){2}
    \Vertex(35,7){4.001}
    \Text(2,12)[lb]{\scalebox{0.8}{$2\Delta_1$}}
    \Text(57,12)[lb]{\scalebox{0.8}{$2\Delta_2$}}
    \Text(101,4)[lb]{$=$}
    \Vertex(125,7){2}
    \Line[arrow,arrowpos=0.5,arrowlength=5,arrowwidth=2,arrowinset=0.2](125,7)(210,7)
    \Vertex(210,7){2}
    \Text(140,12)[lb]{\scalebox{0.8}{$2(\Delta_1+\Delta_2)-d$}}
    \Text(215,2)[lb]{\scalebox{1}{$~\times ~ \pi^\frac{d}{2}\,A(\Delta_2)V(\Delta_1,d-\Delta_1-\Delta_2)$}}
  \end{picture}
  }
\end{center}
\begin{center}
\scalebox{0.9}{
  \begin{picture}(271,37) (49,-10)
    \SetWidth{1.0}
    \SetColor{Black}
    \Vertex(-20,7){2}
    \Line[arrow,arrowpos=0.5,arrowlength=5,arrowwidth=2,arrowinset=0.2](-20,7)(35,7)
    \Line[arrow,arrowpos=0.5,arrowlength=5,arrowwidth=2,arrowinset=0.2](35,7)(85,7)
    \Vertex(85,7){2.001}
    \Vertex(35,7){4}
    \Text(2,12)[lb]{\scalebox{0.8}{$2\Delta_1$}}
    \Text(57,12)[lb]{\scalebox{0.8}{$2\Delta_2$}}
    \Text(101,4)[lb]{$=$}
    \Vertex(125,7){2.001}
    \Line[](125,7)(210,7)
    \Vertex(210,7){2}
    \Text(140,12)[lb]{\scalebox{0.8}{$2(\Delta_1+\Delta_2)-d$}}
    \Text(215,4)[lb]{\scalebox{1}{$~\times~ (-\pi^\frac{d}{2})\,A(d-\Delta_1-\Delta_2) V(\Delta_1,\Delta_2)\times \mathbb{I}$}}
  \end{picture}
  }
\end{center}
Here we defined
\begin{align}
V(\Delta_1,\Delta_2) \,=\, \frac{\Gamma\left(\frac{d}{2}-\Delta_1+\frac{1}{2}\right)}{\Gamma(\Delta_1+\frac{1}{2})}
\frac{\Gamma\left(\frac{d}{2}-\Delta_2+\frac{1}{2}\right)}{\Gamma(\Delta_2+\frac{1}{2})}\,.
\end{align}

Uniqueness relation, valid for $a_1+a_2+a_3=d$, has the form
\begin{equation}
\label{uniqueness}
\int d^dx\,\frac{1}{|x_1-x|^{2a_1} |x_2-x|^{2a_2} |x_3-x|^{2a_3}} 
\,=\,\frac{U(a_1,a_2,a_3)}{|x_{12}|^{d-2a_3}|x_{13}|^{d-2a_2}|x_{23}|^{d-2a_1}}\,,
\end{equation}
and can be represented graphically as
\begin{center}
  \begin{picture}(210,66) (70,-31)
    \SetWidth{1.0}
    \SetColor{Black}
    \Line[](32,34)(80,2)
    \Line[](80,2)(32,-30)
    \Line[](80,2)(128,2)
    \Line[](192,34)(192,-30)
    \Line[](192,-30)(240,2)
    \Line[](240,2)(192,34)
    \Vertex(32,34){2.001}
    \Vertex(32,-30){2.001}
    \Vertex(80,2){4.001}
    \Vertex(128,2){2.00001}
    \Vertex(192,34){2.001}
    \Vertex(192,-30){2.001}
    \Vertex(240,2){2.00001}
    \Text(55,-25)[lb]{\scalebox{0.801}{$2a_1$}}
    \Text(55,22)[lb]{\scalebox{0.801}{$2a_2$}}
    \Text(100,5)[lb]{\scalebox{0.801}{$2a_3$}}
    \Text(160,-1)[lb]{$=$}
    \Text(180,-1)[lb]{\scalebox{0.8001}{$\alpha$}}
    \Text(215,-28)[lb]{\scalebox{0.801}{$\beta$}}
    \Text(215,24)[lb]{\scalebox{0.801}{$\gamma$}}
    \Text(250,-5)[lb]{$\times \left(-\frac{2}{\sqrt{N}}\right) U\left(a_1,a_2,a_3\right)$}
  \end{picture}
\end{center}
Here we have defined $\alpha = d-2a_3$, $\beta = d-2a_2$, $\gamma = d-2a_1$.

In the case of a Yukawa vertex, the uniqueness relation becomes (for $\Delta_1+\Delta_2+\Delta_3 = d$) 
\begin{equation}
 \label{star-triangle}
  \int d^dx_4\,\frac{\gamma_\mu x_{41}^\mu \gamma_\nu x_{24}^\nu}
  {|x_{14}|^{2\Delta_1+1}|x_{24}|^{2\Delta_2+1}|x_{34}|^{2\Delta_3}}\,=\, \frac{\pi^\frac{d}{2}A(\Delta_3)V(\Delta_1,\Delta_2) ~\gamma_\mu x_{31}^\mu
\gamma_\nu x_{23}^\nu}{|x_{12}|^{d-2\Delta_3}|x_{13}|^{d-2\Delta_2+1}|x_{23}|^{d-2\Delta_1+1}} \,,
\end{equation}
or diagrammatically, 
\begin{center}
\scalebox{0.8}{
  \begin{picture}(390,106) (29,-27)
    \SetWidth{1.0}
    \SetColor{Black}
    \Line[arrow,arrowpos=0.5,arrowlength=5,arrowwidth=2,arrowinset=0.2](96,28)(144,76)
    \Line[arrow,arrowpos=0.5,arrowlength=5,arrowwidth=2,arrowinset=0.2](144,-26)(96,28)
    \Line[](96,28)(30,28)
    \Vertex(96,28){4.001}
    \Line[arrow,arrowpos=0.5,arrowlength=5,arrowwidth=2,arrowinset=0.2](196,28)(310,76)
    \Line[arrow,arrowpos=0.5,arrowlength=5,arrowwidth=2,arrowinset=0.2](310,-26)(196,28)
    \Line[](310,76)(310,-26)
    \Text(166,26)[lb]{\scalebox{1}{$=$}}
    \Text(324,22)[lb]{\scalebox{1}{$\times ~ \pi^\frac{d}{2}A(\Delta_3)V(\Delta_1,\Delta_2)$}}
    \Text(60,32)[lb]{\scalebox{0.8}{$2\Delta_3$}}
    \Text(103,58)[lb]{\scalebox{0.8}{$2\Delta_2$}}
    \Text(103,-8)[lb]{\scalebox{0.8}{$2\Delta_1$}}
    \Text(233,60)[lb]{\scalebox{0.8}{$d-2\Delta_1$}}
    \Text(233,-18)[lb]{\scalebox{0.8}{$d-2\Delta_2$}}
    \Text(275,22)[lb]{\scalebox{0.8001}{$d-2\Delta_3$}}
  \end{picture}
  }
\end{center}

\end{document}